\documentclass[%
 reprint,
 superscriptaddress,
 nofootinbib,
 nobibnotes,
 amsmath,amssymb,
 aps,
 prd,
]{revtex4-2}

\usepackage{graphicx}
\usepackage{dcolumn}
\usepackage{bm}
\usepackage[colorlinks=true,linkcolor=blue,citecolor=blue]{hyperref}
\usepackage{xcolor}
\usepackage{mathrsfs}
\usepackage{xfrac}
\usepackage{nicefrac}
\usepackage[caption=false]{subfig}

\begin{document}

\preprint{APS/123-QED}

\title{Electromagnetic responses induced by one- and two-body nuclear currents in inclusive electron-nucleus scattering}

\author{Kajetan Niewczas}
\email{kajetan.niewczas@ugent.be}
\affiliation{École polytechnique, IN2P3-CNRS, Laboratoire Leprince-Ringuet, F-91120 Palaiseau, France}
\affiliation{Department of Physics and Astronomy, Ghent University, Proeftuinstraat 86, B-9000 Gent, Belgium}
\author{Ashish Kumar Jha}
\affiliation{Department of Physics and Astronomy, Ghent University, Proeftuinstraat 86, B-9000 Gent, Belgium}
\author{Raúl González-Jiménez}
\affiliation{Departamento de Física Atómica, Molecular y Nuclear, Universidad de Sevilla, 41080 Sevilla, Spain}
\author{Natalie Jachowicz}
\email{natalie.jachowicz@ugent.be}
\affiliation{Department of Physics and Astronomy, Ghent University, Proeftuinstraat 86, B-9000 Gent, Belgium}
\author{Alexis Nikolakopoulos}
\affiliation{Department of Physics, University of Washington, Seattle, Washington 98195, USA}
\affiliation{Department of Physics and Astronomy, Ghent University, Proeftuinstraat 86, B-9000 Gent, Belgium}
\author{Jan Ryckebusch}
\affiliation{Department of Physics and Astronomy, Ghent University, Proeftuinstraat 86, B-9000 Gent, Belgium}

\date{\today}

\begin{abstract}
    The longitudinal and transverse nuclear response functions encode information about the electromagnetic behavior of the nuclear target probed in inclusive electron-nucleus scattering.
    We present a microscopic calculation based on an independent-particle model, where a non-relativistic nuclear mean-field potential governs both the initial and final hadronic states of the interaction.
    We add dynamically generated short-range nucleon-nucleon correlations (SRCs), as well as meson-exchange currents (MECs) derived from one-pion exchange and intermediate $\Delta$-resonance excitation.
    We evaluate the contributions from one- and two-nucleon knock-out reactions to nuclear responses in inclusive electron scattering from $^{12}\mathrm{C}$ in the quasielastic and dip regions.
    We demonstrate that for quasielastic kinematics, SRCs quench both nuclear responses, while the MECs enhance the transverse signal, substantially improving the predicted transverse-to-longitudinal ratio.
    In the dip region between the quasielastic and the $\Delta$-resonance peaks, we find that the observed excess of transverse strength originates mainly from explicit two-nucleon knock-out following a $\Delta$-resonance excitation.
\end{abstract}

\maketitle


\section{Introduction}
\label{sec:introduction}

The scattering of leptonic probes off atomic nuclei has long been a subject of interest in nuclear physics and related fields. While the specific research motivations have evolved since the pioneering work of N.F.~Mott~\cite{Mott:1929}, the quest to refine our understanding of nuclear dynamics continues to stimulate further studies. Over the past two decades, renewed interest in this topic has been driven largely by accelerator-based neutrino experiment programs~\cite{NuSTEC:2017hzk}. At the GeV-scale energies explored in these experiments, modeling neutrino-nucleus interactions remains a substantial source of systematic uncertainties, particularly when measuring the parameters of neutrino oscillations~\cite{T2K:2011qtm,NOvA:2016vij}. Uncovering a potentially non-zero value of the CP-violating phase in the leptonic sector at a statistically meaningful level~\cite{T2K:2019bcf,Hyper-Kamiokande:2018ofw,DUNE:2020lwj} will require a new level of precision in both cross section and oscillation measurements, which in turn demands deeper insights into nuclear structure and improved methods for modeling neutrino-nucleus scattering. Due to the correspondence between the underlying physics of electron and neutrino scattering, along with similarities in the nuclear responses, an accurate comparison with electron scattering data has been a benchmark for the quality of lepton-nucleus scattering models~\cite{Amaro:2019zos,Lovato:2023raf}.

Describing electron-nucleus interactions in the intermediate-energy region requires bridging fundamental particle physics with complex nuclear structure. While electrons are considered elementary spin-$\sfrac{1}{2}$ Dirac particles, nuclei are strongly interacting many-body systems requiring phenomenological modeling. Upon transferring hundreds of $\mathrm{MeV}$ to a nucleus---more than the average binding energy per nucleon but less than the internal binding of quarks---one can describe the nucleus, to a reasonable degree of accuracy, as a collection of quasifree nucleons~\cite{Walecka:1995mi}. Examples of \textit{independent-particle models} (IPMs) are a Fermi gas (FG) of nucleons or a Slater determinant of single-particle states in a nuclear mean-field potential obtained in the Hartree-Fock (HF) framework. Alternatively, one can use the realistic hole spectral function (SF) formalism and extend the latter with the effect of nucleon-nucleon correlations evaluated in the local-density approximation~\cite{Benhar:1994hw}, or make the mean-field calculation relativistic (RMF)~\cite{Gonzalez-Jimenez:2019ejf}. By treating the electromagnetic interaction as a single-boson exchange in the \textit{impulse approximation} (IA)~\cite{Frullani:1984nn}, these models can reasonably describe the main features of inclusive $(e,e^\prime)$ scattering cross sections~\cite{Moniz:1971mt,Donnelly:1970rk}. However, this approach does have limitations, as an accurate and simultaneous description of the cross section and the separated longitudinal ($W_{L}$) and transverse ($W_{T}$) nuclear response functions requires more sophisticated theoretical modeling. As summarized in Refs.~\cite{Boffi:1996ikg,Benhar:2006wy}, finding and understanding adequate mechanisms has been a long-standing challenge.

Calculations based on IPMs involving only one-body nuclear currents tend to underestimate the transverse-to-longitudinal response ratio compared to Rosenbluth-type analyses of inclusive electron-nucleus cross sections~\cite{Barreau:1983ht,Jourdan:1996np}. At first glance, this discrepancy comes mainly from the overestimation of the longitudinal strength, which, historically, led to the investigation of the Coulomb sum rule (CSR) and the effect of short-range nucleon-nucleon correlations (SRCs) therein~\cite{CZYZ196347}. Many nuclear mechanisms have been proposed to quench the longitudinal response, such as medium-modified nucleon properties~\cite{Noble:1980my,Celenza:1986vq,Mulders:1987ka}, final-state interactions (FSI)~\cite{Chinn:1988ky,Boffi:1991hx,Capuzzi:1991qd}, collective effects, such as those captured in the random phase approximation (RPA)~\cite{Alberico:1987zz,Brieva:1987zz,Co:1988zku,MARIASARUIS199357,Bauer:2003yv,Pandey:2014tza,Jachowicz:2021ieb}, and relativistic effects~\cite{DeForest:1983ahx,Wehrberger:1987ri,Horowitz:1990bm}. Most of these corrections comparably impact the responses, thus failing to account for the observed transverse-to-longitudinal ratio. As suggested e.g. in Ref.~\cite{VanderSluys:1995rp}, the solution to this challenge emerges from introducing two-body nuclear currents with a virtual meson exchanged between two nucleons. In electromagnetic interactions, these meson-exchange currents (MECs) contribute primarily to the transverse response, leading to one or more nucleons being knocked out of the nucleus. The necessity of including MECs to provide the required transverse enhancement has been robustly demonstrated by ab-initio many-body calculations~\cite{Lovato:2016gkq,Pastore:2019urn}.

Meson-exchange currents, first proposed to explain the discrepancy between the magnetic moments of the $^{3}\mathrm{H}$ and $^{3}\mathrm{He}$ nuclei~\cite{Villars:1947}, have been extensively investigated in the context of photoabsorption and electron scattering. At the quasielastic (QE) peak in inclusive electron scattering, their contribution to one-nucleon knock-out is relatively small; however, the interference between the one- and two-body currents significantly modifies the transverse response and visibly alters the computed cross sections. Many theoretical methods have been developed to study this effect, finding differences even in the relative signs of particular MEC contributions. In the non-relativistic formulation, the net interference between the dominant one-body and the two-body currents was found to be destructive in FG models~\cite{Kohno:1981dg,Fabrocini:1996bu}, destructive in analyses with both FG and HF frameworks~\cite{Alberico:1989aja,Amaro:1994fx}, or constructive in HF~\cite{VanderSluys:1995rp}. Meanwhile, relativistic formulations of meson-exchange currents found a constructive net interference in FG~\cite{Dekker:1992px}, RMF~\cite{Franco-Munoz:2023zoa,Franco-Munoz:2022jcl}, and SF frameworks~\cite{Lovato:2023khk}, but a destructive effect in other FG calculations~\cite{Amaro:2001xz,Amaro:2003yd,Casale:2023hyz}. The recent ab-initio Green's function Monte Carlo~\cite{Carlson:2001mp,Lovato:2016gkq} and the short-time approximation~\cite{Pastore:2019urn} calculations of inclusive electromagnetic responses find a constructive net effect of including non-relativistic MECs. However, these calculations employ fundamentally different many-body nuclear states that go beyond the IPM by explicitly incorporating nucleon-nucleon correlations; thus, a direct comparison with other models is not straightforward. At present, Refs.~\cite{Lovato:2023khk,Franco-Munoz:2026hxi} attribute discrepancies in the qualitative behavior of the interference between one- and two-body currents to the relative sign of the $\Delta$-current contribution, which remains theoretically ambiguous.

In the \textit{dip region} between the QE and $\Delta$-resonance peaks, the transverse response dominates the Rosenbluth-decomposed inclusive electron scattering data~\cite{Barreau:1983ht,Jourdan:1996np}. Accounting for this excess requires the involvement of explicit pion degrees of freedom~\cite{Towner:1987zz}. This is achieved by evaluating MEC diagrams, which contain virtual-pion lines and thus lead to two-nucleon knock-out final states. These contributions to the transverse response in inclusive electron scattering have been investigated in several calculations. Non-relativistic currents were evaluated within FG~\cite{Alberico:1983zg,Martini:2011ui} and HF~\cite{Amaro:1994fx,VanderSluys:1995rp} frameworks, with Ref.~\cite{VanderSluys:1995rp} extending the non-relativistic limitation of the static $\Delta$-resonance description via the \textit{resonating} approximation of Ref.~\cite{Dekker:1994yc}. Relativistic meson-exchange currents were calculated within FG~\cite{Dekker:1991ph,Gil:1997bm,DePace:2003spn,RuizSimo:2016rtu,Megias:2016lke,Belocchi:2025lqp} and SF~\cite{Benhar:2015ula} frameworks. As predicted by the pioneering work of T.W.~Donnelly \textit{et al.}~\cite{Donnelly:1978xa}, which endorsed response-separated inclusive electron scattering data as a powerful tool to study two-body nuclear dynamics, all above-mentioned results agree that two-nucleon knock-out provides the necessary strength to reproduce data up to the $\Delta$-resonance peak.

The Ghent group has developed a theoretical framework for intermediate-energy electron- and neutrino-nucleus scattering built upon a non-relativistic HF nuclear mean field. In our previous works, we incorporated several physical mechanisms to refine this description, including long-range nuclear correlations via continuum RPA (CRPA)~\cite{Ryckebusch:1988bpw,Jachowicz:2002rr,Pandey:2014tza,Jachowicz:2021ieb}, short-range nucleon-nucleon correlations~\cite{Ryckebusch:1997gn,Janssen:1999xy,VanCuyck:2016fab}, and meson-exchange currents~\cite{Ryckebusch:1993tf,VanderSluys:1995rp,Ryckebusch:1999gu,VanCuyck:2017wfn}. In this work, we present numerical calculations that simultaneously incorporate both SRC and MEC dynamics. Specifically, we evaluate the longitudinal and transverse nuclear response functions, as well as their ratio, in inclusive electron scattering on $^{12}\mathrm{C}$. We benchmark our calculations against the measured energy dependence of the $^{12}\mathrm{C}(e,e')$ cross sections, focusing primarily on the three kinematics reported by Ref.~\cite{Barreau:1983ht}, with additional comparisons to the data of Ref.~\cite{Jourdan:1996np}.

This paper is organized as follows. In Sec.~\ref{sec:framework}, we review the kinematics and cross section formalism for electron-induced one- and two-nucleon knock-out reactions and describe the foundations of our nuclear framework. In Sec.~\ref{sec:interaction}, we present the dynamics of the investigated interactions and describe the essential components of the model. In Sec.~\ref{sec:results}, we report the results of our study and provide a detailed discussion. Finally, in Sec.~\ref{sec:conclusions}, we present our conclusions.

\section{Theoretical framework}
\label{sec:framework}

\subsection{Kinematics and cross section}

\begin{figure}[t]
  \centering
  \includegraphics[width=0.48\textwidth]{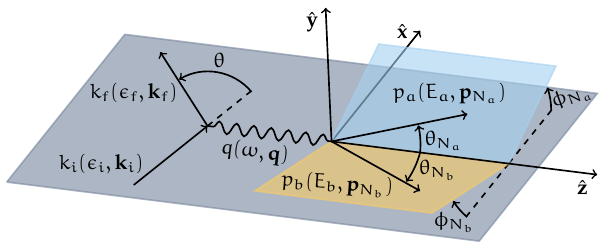}
  \caption{Kinematics of the two-nucleon knock-out process. The $z$-axis coincides with the direction of momentum transfer $\mathbf{q}$. The nucleon (reaction) kinematics are defined relative to the lepton scattering $xz$-plane.}
  \label{fig:kinematics}
\end{figure}

Capturing the full dynamics of intermediate-energy electrons scattering off atomic nuclei is a complex multidimensional problem engaging several approximations and advanced modeling techniques. We employ the Born picture, where an incoming lepton with four-momentum $k_i = (\epsilon_i,\mathbf{k}_i)$, exchanges a single virtual boson with the hadronic system to a final observed lepton state $k_f = (\epsilon_f,\mathbf{k}_f)$. In this work, we investigate the circumstances under which this interaction leads to the following processes: the electron-induced one-nucleon knock-out
\begin{equation}
\label{eq:1p1h}
    e + A \to e^\prime + (A-1)^\ast + N_a,
\end{equation}
and two-nucleon knock-out
\begin{equation}
\label{eq:2p2h}
    e + A \to e^\prime + (A-2)^\ast + N_a + N_b.
\end{equation}
We denote the initial and final hadronic states as $P_A = (E_A,\mathbf{P}_A)$ and $P_R = (E_R,\mathbf{P}_R)$, respectively. The residual nucleus $R$ corresponds to $(A-1)^\ast$ or $(A-2)^\ast$. The asterisk indicates an excited nuclear remnant, which remains unstable with an excitation energy reaching tens of~$\mathrm{MeV}$. The four-momenta of the outgoing nucleon(s) are $p_{N_a} = (E_{N_a},\mathbf{p}_{N_a})$ and $p_{N_b} = (E_{N_b},\mathbf{p}_{N_b})$. Working in the laboratory frame, the target nucleus is initially at rest, $P_A = (M_A,\mathbf{0})$. We define the four-momentum transfer $q = (\omega,\mathbf{q})$ through
\begin{equation}
    \omega = \epsilon_i - \epsilon_f, \qquad \mathbf{q} = \mathbf{k}_i - \mathbf{k}_f,
\end{equation}
and the invariant quantity $Q^2$ as $-q^2 = -(k_i - k_f)^2$. As presented for two-nucleon knock-out in Fig.~\ref{fig:kinematics}, we work in the coordinate system with $\mathbf{q}$ along the $z$-axis and the (electron) scattering plane coinciding with the $xz$-plane.

The inclusive ($e,e^\prime$) differential cross section is expressed as
\begin{widetext}
\begin{equation}
    \frac{\mathrm{d}^3 \sigma (e,e^\prime)}{\mathrm{d}\epsilon_f \mathrm{d}\Omega_f} = \sigma^\mathrm{Mott} \left\{ \left( \frac{Q^2}{|\mathbf{q}|^2} \right)^2 W_{L}(\omega,\mathbf{q}) + \left( \frac{Q^2}{2|\mathbf{q}|^2} + \tan^2(\theta/2) \right) W_{T}(\omega,\mathbf{q}) \right\},
\end{equation}
where $\sigma^\mathrm{Mott} = \frac{\alpha^2 \cos^2(\theta/2)}{4 \epsilon_i^2 \sin^4(\theta/2)}$ is the Mott cross section, which describes scattering off a point-like particle, and $\alpha = e^2/4\pi$ is the fine-structure constant. The $W_L$ and $W_T$ encode the electromagnetic response of the nucleus along and about the direction of the momentum transfer. We construct them from bilinear products of nuclear transition matrix elements of the electromagnetic current operator $\hat{J}_\mu = \left(\hat{\rho},\hat{\bm{J}}\right) = \left(\hat{\rho},\hat{J}_x,\hat{J}_y,\hat{J}_z\right)$ as
\begin{align}
\label{eq:response:long}
    W_L (\omega,\mathbf{q}) & = \sum_B \left| \langle \, \Psi_B \, | \, \hat{\rho}(\mathbf{q}) \, | \, \Psi_A \, \rangle \right|^2 \delta(\omega - E_B + E_A), \\
\label{eq:response:trans}
    W_T (\omega,\mathbf{q}) & = \sum_B \left\{ \left| \langle \, \Psi_B \, | \, \hat{J}_{+1}(\mathbf{q}) \, | \, \Psi_A \, \rangle \right|^2 + \left| \langle \, \Psi_B \, | \, \hat{J}_{-1}(\mathbf{q}) \, | \, \Psi_A \, \rangle \right|^2 \right\} \delta(\omega - E_B + E_A),
\end{align}
where the sum over $B$ extends over all possible final hadronic states, with an excitation energy $E_B-E_A$, relative to the ground-state energy $E_A$ of the target nucleus $| \, \Psi_A \, \rangle$ ($J^\pi = 0^+$). For the one- (Eq.~\eqref{eq:1p1h}) and two-nucleon (Eq.~\eqref{eq:2p2h}) knock-out reactions investigated in this work, the summation over the final state $| \, \Psi_B \, \rangle$ runs over all possible \textit{1-particle-1-hole} (1p1h) and \textit{2-particle-2-hole} (2p2h) excitations of the ground state $| \, \Psi_A \, \rangle$. Therefore, we express the inclusive cross section formula as
\begin{equation}
    \frac{\mathrm{d}^3 \sigma (e,e^\prime)}{\mathrm{d}\epsilon_f \mathrm{d}\Omega_f} = \sum_{N_a} \int \mathrm{d}\Omega_{N_a} \mathrm{d}T_{N_a} \frac{\mathrm{d}^6 \sigma (e,e^\prime N_a)}{\mathrm{d}\epsilon_f \mathrm{d}\Omega_f \mathrm{d}\Omega_{N_a} \mathrm{d}T_{N_a}} + \sum_{N_a,N_b} \int \mathrm{d}\Omega_{N_a} \mathrm{d}\Omega_{N_b} \mathrm{d}T_{N_a} \mathrm{d}T_{N_b} \frac{\mathrm{d}^9 \sigma (e,e^\prime N_a N_b)}{\mathrm{d}\epsilon_f \mathrm{d}\Omega_f \mathrm{d}\Omega_{N_a} \mathrm{d}T_{N_a} \mathrm{d}\Omega_{N_b} \mathrm{d}T_{N_b}},
\end{equation}
\end{widetext}
which includes integrals over hadronic degrees of freedom for each final state. For the single-nucleon knock-out, the kinetic energy of the outgoing nucleon $T_{N_a}$ is constrained by the energy conservation relation $\omega = E^\ast_{A-1} + S_N + T_{N_a} + T_{A-1}$, where $E^\ast_{A-1}$ is the excitation energy of the residual $(A-1)$ nucleus, and $S_N$ is the one-nucleon separation energy. For two-nucleon knock-out, energy conservation leads to $\omega = E^\ast_{A-2} + S_{2N} + T_{N_a} + T_{N_b} + T_{A-2}$, where $E^\ast_{A-2}$ is the excitation energy of the residual $(A-2)$ nucleus, and $S_{2N}$ is the two-nucleon separation energy.

\subsection{Mean-field approach}

We employ an independent-particle framework, where both bound and scattering nucleon states are obtained by solving the Schr{\"o}dinger equation for a mean-field potential incorporating central, spin-orbit, and Coulomb terms. This potential is derived via an iterative Hartree-Fock calculation using an effective Skyrme-type nucleon-nucleon force, specifically the SkE2 parameterization~\cite{Waroquier:1983fqb}, a well-established framework that aims to reflect the ground-state and low-lying excited state characteristics of spherical nuclei~\cite{Vautherin:1971aw}. While the model requires no empirical input for the initial or final-state potentials, we enhance its accuracy by aligning bound-state energy levels with experimental single-nucleon separation energies, ensuring closer agreement with observed energy spectra~\cite{Jachowicz:2000phd}.

\begin{figure}[t]
  \centering
  \includegraphics[width=0.37\textwidth]{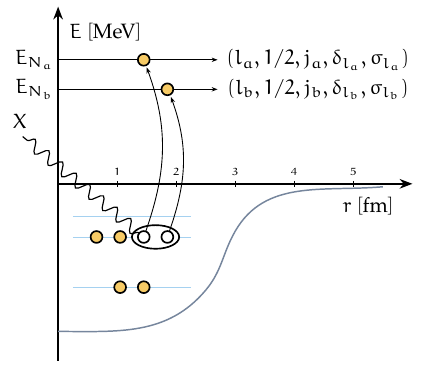}
  \caption{Two-nucleon knock-out process, where the interacting boson ejects two nucleons to the continuum, and the residual nucleus remains in a two-hole state: $(hh')^{-1}$.}
  \label{fig:shell_model}
\end{figure}

As illustrated for the two-nucleon knock-out case in Fig.~\ref{fig:shell_model}, the ejected nucleons are asymptotically free, yet they are subject to the influence of the mean-field potential of the target nucleus. This represents a distorted-wave calculation in the so-called \textit{spectator approach}. Because both bound and final scattering states are generated within the same potential, they are strictly orthogonal, inherently satisfying the Pauli exclusion principle without requiring manual phase-space cuts. Recoil of the residual nucleus is neglected, and no inelastic final-state interactions between the outgoing nucleons and the residual system are included. These approximations make the calculation computationally feasible while retaining a quantum-mechanical treatment of the final state.

Evaluating the hadronic responses of Eqs.~\eqref{eq:response:long} and~\eqref{eq:response:trans}, requires the computation of specific nuclear transition matrix elements. In our framework, reactions with one or two nucleons in the continuum lead to the following transitions:
\begin{widetext}
\begin{equation}
    J_\mu (\mathbf{q}) = \langle \, \Psi_B \, | \, \hat{J}_{\mu}(\mathbf{q}) \, | \, \Psi_A \, \rangle \simeq
    \sum_{\mathrm{1p1h}}
    \langle \, \Phi_{B} \, | \, \Phi_{\mathrm{1p1h}} \, \rangle \langle \, \Phi_{\mathrm{1p1h}} \, | \, \hat{J}_{\mu}(\mathbf{q}) \, | \, \Phi_A \, \rangle +
    \sum_{\mathrm{2p2h}}
    \langle \, \Phi_{B} \, | \, \Phi_{\mathrm{2p2h}} \, \rangle \langle \, \Phi_{\mathrm{2p2h}} \, | \, \hat{J}_{\mu}(\mathbf{q}) \, | \, \Phi_A \, \rangle ,
\end{equation}
where the many-body wave functions $\left( | \, \Psi_A \, \rangle \right.$, $\left.| \, \Psi_B \, \rangle \right)$ are approximated by Slater determinants $\left( | \, \Phi_A \, \rangle \right.$, $\left.| \, \Phi_B \, \rangle \right)$, ensuring that contributions from distinct final states add incoherently. After performing a partial-wave expansion in terms of continuum eigenstates for each outgoing nucleon, we obtain~\cite{Ryckebusch:1993tf}:
\begin{equation}
\begin{split}
\label{eq:1p1h_state}
    | \, \Phi_{1p1h} \, \rangle
    = & | \, \Phi_{A-1} (E_\mathrm{exc}^{A-1}, J_R M_R); \, \mathbf{p}_{N_a} m_{s_{N_a}} \, \rangle_\mathrm{as} \\
    = & \sum_{\alpha_h,m_{j_h}} \sum_{{l_a},m_{l_a}} \sum_{{j_a},m_{j_a}} 4\pi \frac{\sqrt{\pi}}{\sqrt{2M_N|\mathbf{p}_{N_a}|}} i^{{l_a}} e^{i(\delta_{l_a} + \sigma_{l_a})} Y_{{l_a},m_{l_a}}^\ast (\Omega_{N_a}) (-1)^{j_h+m_{j_h}} \\
    & \times \delta_{j_h,J_R} \delta_{m_{j_h},M_R} \langle \, {l_a}, m_{l_a}; \, \sfrac{1}{2}, m_{s_{N_a}} \, | \, {j_a}, m_{j_a} \, \rangle \widehat{c}^{\, \dagger}_{l_a,j_a,m_{j_a}} \widehat{c}_{\alpha_h,-m_{j_h}} | \, \Phi_A \, \rangle ,
\end{split}
\end{equation}
and
\begin{equation}
\begin{split}
\label{eq:2p2h_state}
    | \, \Phi_{2p2h} \, \rangle
    = & | \, \Phi_{A-2} (E_\mathrm{exc}^{A-2}, J_R M_R); \, \mathbf{p}_{N_a} m_{s_{N_a}}; \, \mathbf{p}_{N_b} m_{s_{N_b}} \, \rangle_\mathrm{as} \\
    = & \sum_{\alpha_h,m_{j_h}} \sum_{\alpha_{h'},m_{j_{h'}}} \sum_{{l_a},m_{l_a}} \sum_{{j_a},m_{j_a}} \sum_{{l_b},m_{l_b}} \sum_{{j_b},m_{j_b}} \frac{1}{\sqrt{1+\delta_{h,{h'}}}} (4\pi)^2 \frac{\pi}{2M_N\sqrt{|\mathbf{p}_{N_a}||\mathbf{p}_{N_b}|}} \\
    & \times i^{{l_a}+{l_b}} e^{i(\delta_{l_a} + \sigma_{l_a}+\delta_{l_b} + \sigma_{l_b})} Y_{{l_a},m_{l_a}}^\ast (\Omega_{N_a}) Y_{{l_b},m_{l_b}}^\ast (\Omega_{N_b}) (-1)^{j_h+m_{j_h}+j_{h'}+m_{j_{h'}}} \\
    & \times \langle \, j_h, m_{j_h}; \, j_{h'}, m_{j_{h'}} \, | \, J_R, M_R \, \rangle \langle \, {l_a}, m_{l_a}; \, \sfrac{1}{2}, m_{s_{N_a}} \, | \, {j_a}, m_{j_a} \, \rangle \\
    & \times \langle \, {l_b}, m_{l_b}; \, \sfrac{1}{2}, m_{s_{N_b}} \, | \, {j_b}, m_{j_b} \, \rangle \widehat{c}^{\, \dagger}_{l_a,j_a,m_{j_a}} \widehat{c}^{\, \dagger}_{l_b,j_b,m_{j_b}} \widehat{c}_{\alpha_h,-m_{j_h}} \widehat{c}_{\alpha_{h'},-m_{j_{h'}}} | \, \Phi_A \, \rangle ,
\end{split}
\end{equation}
where $|\cdot\rangle_\mathrm{as}$ denotes a fully antisymmetrized state. The second quantization operators, $\widehat{c}^{\, \dagger}_{l,j,m}$ and $\widehat{c}_{l,j,m}$, allow the creation or annihilation of single-nucleon states, which, in our spherically-symmetric basis, are of the form $\psi_{n,l,\sfrac{1}{2},j,m} (|\mathbf{r}|) = \phi_{n,l,j}(|\mathbf{r}|) \mathcal{Y}^m_{j,(l,\sfrac{1}{2})}(\Omega_r)$, where $\phi_{n,l,j}(|\mathbf{r}|)$ are the radial wave functions, and $\mathcal{Y}^m_{j,(l,\sfrac{1}{2})}(\Omega_r)$ are the vector spherical harmonics. The bound nucleon states are uniquely identified by the set of quantum numbers $\alpha = (n,l,\sfrac{1}{2},j)$. The outgoing nucleon states have the following asymptotic behavior~\cite{Ryckebusch:1988bpw}:
\begin{equation}
\label{eq:asymptotic}
    \phi_{l,j}(|\mathbf{r}|) \xrightarrow[]{|\mathbf{r}| \gg |\mathbf{r}_A|} \ \sqrt{\frac{2M_N|\mathbf{p}_N|}{\pi}} \frac{\sin(|\mathbf{p}_N| |\mathbf{r}| - \eta \ln(2|\mathbf{p}_N| |\mathbf{r}|) - \pi l / 2 + \delta_l + \sigma_l)}{|\mathbf{p}_N| |\mathbf{r}|} ,
\end{equation}
\end{widetext}
where $|\mathbf{r}_A|$ is the nuclear radius, $\delta_l$ and $\sigma_l$ are the central and Coulomb phase shifts, and the factor $\eta$ accounts for the Coulomb part of the single-particle potential. For clarity of notation, we omit explicit references to the isospin dependence of nucleon states, but note that for neutrons, $\sigma_l = \eta = 0$.

Hadronic final states constructed in this manner satisfy our model assumptions and are well suited for evaluating nuclear transition matrix elements of electromagnetic current operators. In our numerical calculations, a multipole expansion is adopted for the currents, following the standard expressions for multipole operators (e.g., Ref.~\cite{Walecka:1995mi}). The convergence of this multipole expansion aligns with analogous calculations in the literature~\cite{Amaro:1993306} as well as our previous works~\cite{VanCuyck:2016fab,VanCuyck:2017wfn}, while a comprehensive convergence analysis of the adopted framework can be found in Ref.~\cite{Niewczas:2023phd}.

\section{Interaction dynamics}
\label{sec:interaction}

\subsection{Impulse approximation}
\label{subsec:impulse_approximation}

\begin{figure}[t]
  \centering
  \subfloat[]{\includegraphics[width=0.19\textwidth]{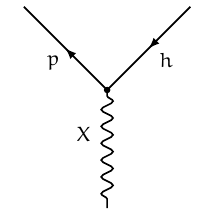}}
  \subfloat[]{\includegraphics[width=0.19\textwidth]{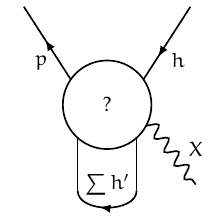}}
  \caption{Graphical representation of the diagrams contributing to the one-nucleon knock-out process considered in this work: (a) one-body current, (b) two-body current.}
  \label{fig:diagrams_1p1h}
\end{figure}

Within the Born approximation, we model the electron-nucleus interaction as an exchange of a single virtual boson, effectively isolating dynamics to individual nucleon degrees of freedom and treating the nuclear current operator as a sum of many-body currents. We truncate this expansion to include one- and two-body operators, i.e., $J_\mu(\mathbf{q}) = J^{[1]}_\mu(\mathbf{q}) + J^{[2]}_\mu(\mathbf{q})$. As illustrated in Fig.~\ref{fig:diagrams_1p1h}, the one-nucleon knock-out reaction may result from interactions on single-nucleon targets or two-nucleon states with only one nucleon emitted to the continuum. The impulse approximation assumes that all other interactions are negligible, effectively retaining solely the one-body nuclear current operator. The latter case, involving two-body currents with a loop over one of the hole states, is addressed in the subsequent subsections.

\begin{figure*}[t]
  \centering
  \includegraphics[width=1.0\textwidth]{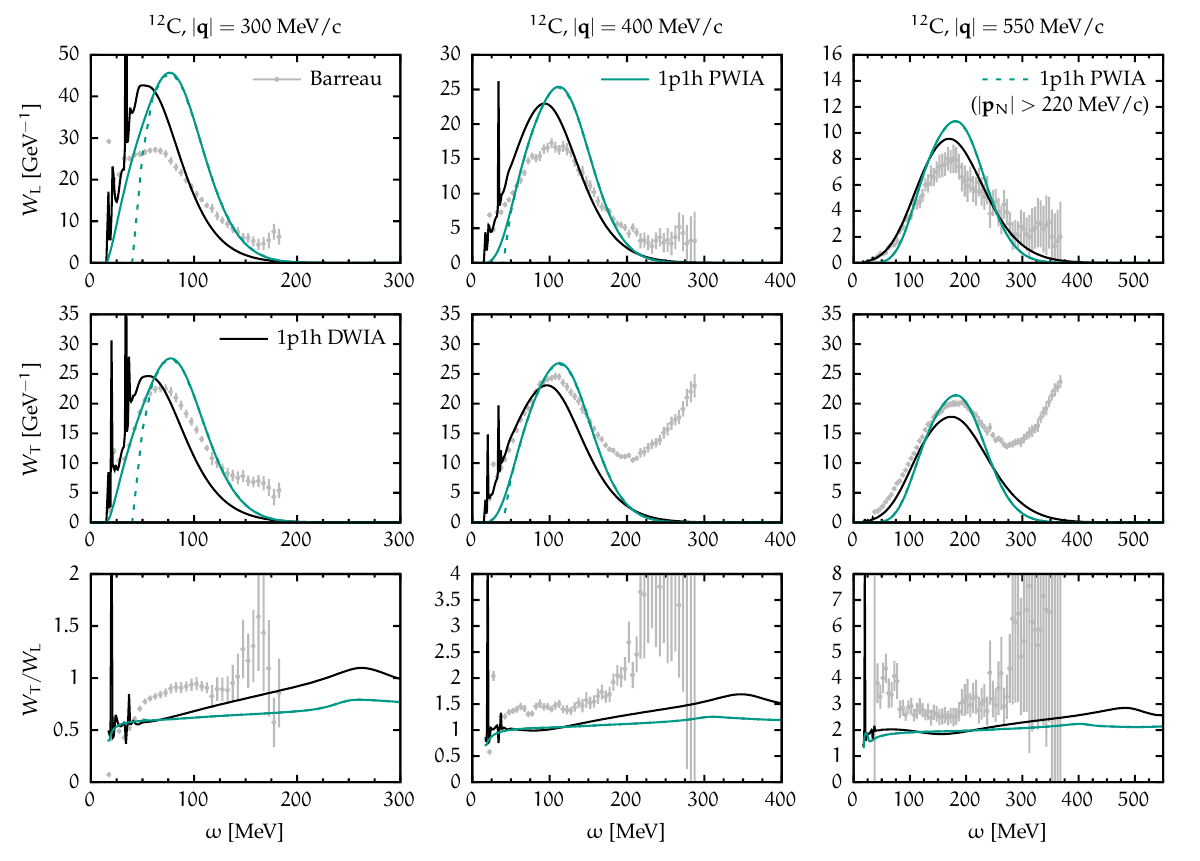}
  \caption{Energy dependence of the 1p1h hadronic $^{12}\mathrm{C}(e,e')$ responses and their ratio for fixed three-momentum transfer values, compared with the experimental results of Ref.~\cite{Barreau:1983ht}. The black solid lines present the impulse approximation calculation, while the teal solid and dashed lines provide the plane-wave impulse approximation results without and with a Fermi momentum cut on the outgoing nucleon, respectively. Note that in the bottom row, both PWIA predictions overlap.}
  \label{fig:resp_pwdw}
\end{figure*}

We derive the one-body currents from the on-shell covariant single-nucleon current~\cite{Bjorken:1965sts}
\begin{equation}
    \hat{J}^{[1]}_\mu (q) = F_1(Q^2) \gamma_\mu
    + \frac{i}{2M_N} F_2(Q^2)\sigma_{\mu\nu} q^\nu ,
\end{equation}
where $F_1(Q^2)$ and $F_2(Q^2)$ are the Dirac and Pauli form factors. Following the Foldy-Wouthuysen non-relativistic reduction, retaining terms up to the order $\mathcal{O}(\sfrac{|\mathbf{q}|}{M_N})$, we obtain in coordinate space
\begin{align}
\label{eq:ia_time}
    \hat{\rho}^{[1],\mathrm{IA}} (\mathbf{r}) & = \sum_{i=1}^A \hat{e}_i G_D(Q^2) \delta^{(3)}(\mathbf{r}_i - \mathbf{r}), \\
\begin{split}
\label{eq:ia_space}
    \hat{\bm{J}}^{\, [1],\mathrm{IA}} (\mathbf{r}) & =
    \sum_{i=1}^A \frac{\hat{e}_i G_D(Q^2)}{2M_N i} \left\{ \bm{\nabla}_i , \delta^{(3)}(\mathbf{r}_i - \mathbf{r}) \right\} \\
    & + \sum_{i=1}^A \frac{\hat{\mu}_i G_D(Q^2)}{2M_N} \bm{\nabla} \times \bm{\sigma}_i \delta^{(3)}(\mathbf{r}_i - \mathbf{r}) ,
\end{split}
\end{align}
where $\{\cdot,\cdot\}$ denotes the anticommutator. The electric charge and magnetic moment operators are
\begin{align}
    \hat{e}_i & = \frac{1}{2} (1 + (\tau_i)_z), \\
    \hat{\mu}_i & = 2.793 \frac{1}{2} (1 + (\tau_i)_z) - 1.913 \frac{1}{2} (1 - (\tau_i)_z) ,
\end{align}
with
\begin{equation}
    \tau_z | \, p \, \rangle = + | \, p \, \rangle, \ \ \ \ \ \tau_z | \, n \, \rangle = - | \, n \, \rangle .
\end{equation}
The expressions in Eqs.~\eqref{eq:ia_time} and~\eqref{eq:ia_space} align with the typically used non-relativistic electromagnetic one-body currents (e.g., Ref.~\cite{Walecka:1995mi}). While these current operators are derived via a non-relativistic expansion, we retain the full $Q^2$-dependence of the form factors rather than their static $|\mathbf{q}|^2$-dependent approximations. This ensures that the hadronic structure is treated consistently with the relativistic kinematics employed throughout this work. The electric and magnetic Sachs form factors are given by $\hat{e}_i G_D(Q^2)$ and $\hat{\mu}_i G_D(Q^2)$, for which we have adopted the standard dipole form
\begin{equation}
    G_D(Q^2) = \frac{1}{(1 + \frac{Q^2}{M_V^2})^{2}},
\end{equation}
with $M_V = 843~\mathrm{MeV}/c^2$. For the neutron electric form factor, we use the Galster parametrization~\cite{Galster:1971kv}:
\begin{equation}
    G_{E,n} (Q^2) = 1.913 \frac{\tau}{1+5.6\tau} G_D(Q^2),
\end{equation}
where $\tau = \sfrac{Q^2}{4M_N^2}$. Note that, in our framework, the summation over bound nucleons $i$ (Eqs.~\eqref{eq:ia_time} and~\eqref{eq:ia_space}) and occupied mean-field nucleon states (Eqs.~\eqref{eq:1p1h_state} and~\eqref{eq:2p2h_state}) are equivalent.

While our non-relativistic framework offers high computational efficiency for distorted-wave calculations of lepton-nucleus interactions, we need to incorporate a series of \textit{relativistic corrections}. First, we take the kinematics of the process, as well as the leptonic part of the interaction, in exact relativistic form. Second, on the hadronic side, we correct the position of the QE peak to its relativistic value. As shown in Ref.~\cite{Alberico:1989aja}, this adjustment is achieved by an effective shift of the energy transfer:
\begin{equation}
    \omega \to \omega \left( 1 + \frac{\omega}{2M_N} \right) ,
\end{equation}
which is applied while determining the energy of outgoing nucleons. Additional relativistic corrections require an expansion of the nuclear currents in higher powers of $(\sfrac{\mathbf{q}}{M_N})$. In Ref.~\cite{Ryckebusch:1999gu}, one employs currents expanded up to order $\mathcal{O}(\sfrac{|\mathbf{q}|^{\, 2}}{M_N^2})$. Incorporating the spin-independent terms is relatively straightforward~\cite{Friar:1973hxj}, but doing so without including the non-trivial spin-orbit term would result in inconsistencies. Implementing spin-orbit current corrections alongside SRCs leads to severe technical complexity in the matrix elements and lies beyond the scope of this study.

In Fig.~\ref{fig:resp_pwdw}, we present the one-body current calculation of electromagnetic responses compared to the experimental results of Ref.~\cite{Barreau:1983ht}. The one-body results overestimate the longitudinal response while underestimating the transverse response---a behavior directly reflected in the underestimation of the $W_T/W_L$ ratio. While a calculation in the \textit{plane-wave impulse approximation} (PWIA), shown for comparison, appears to reproduce the transverse strength reasonably well at higher $|\mathbf{q}|$, it fails to capture the data at low $|\mathbf{q}|$. This shape is much better described by the \textit{distorted-wave impulse approximation} (DWIA), particularly in the low-$\omega$ region below the QE peak, highlighting the role of FSI. Introducing the distortion of the outgoing nucleon via non-vanishing phase shifts ($\sigma_l,\delta_l \neq 0$) in Eq.~\eqref{eq:asymptotic} quenches both responses and shifts the strength towards lower energy transfer values. Scattering into near-threshold continuum energy eigenstates manifests itself as characteristic sharp peaks at low $\omega$, where the lowest multipoles dominate the strength.

We now turn to the treatment of the Pauli exclusion principle. Because the PWIA description inherently lacks Pauli blocking effects, a commonly used ad-hoc fix is to strictly require outgoing nucleons to have a momentum above the Fermi sea (dashed curves in Fig.~\ref{fig:resp_pwdw}). This introduces an unphysical cutoff that artificially truncates the strength, relying on a sharp step-function that is not supported by realistic calculations of nucleon momentum distributions~\cite{Ryckebusch:2019oya}. In the DWIA calculation, Pauli blocking effects emerge from the orthogonality of the wave functions. Nevertheless, DWIA and PWIA calculations consistently underestimate the $W_T/W_L$ ratio, which is a characteristic feature of calculations including solely one-body nuclear currents. We elaborate on our solutions to this issue in the following subsections.

\subsection{Meson-exchange currents}
\label{subsec:mec}

A time-honored view of the medium- and long-range parts of the nucleon-nucleon force is based on the exchange of virtual mesons, such as pions ($\pi$), rho mesons ($\rho$), and omega mesons ($\omega$)~\cite{Ericson:1988gk}. While our HF procedure accounts for dressed nucleon states, the photon-nucleus interaction in the impulse approximation does not explicitly include mesonic degrees of freedom. Meson-exchange currents address this by incorporating two-body mechanisms, i.e., scattering on pairs of nucleons, leading to one or both particles emitted to the continuum. In this work, we focus exclusively on diagrams involving the exchange of the lightest meson, the pion ($m_\pi \simeq 135~\mathrm{MeV}/c^2$). The heavier mesons ($m_\rho \simeq 775~\mathrm{MeV}/c^2$, $m_\omega \simeq 782~\mathrm{MeV}/c^2$) have significantly shorter interaction ranges due to their larger masses. We describe our methodology for including short-range effects in Subsection~\ref{subsec:src}.

\subsubsection{Seagull and pion-in-flight currents}

\begin{figure*}[t]
  \centering
  \subfloat[]{\includegraphics[width=0.19\textwidth]{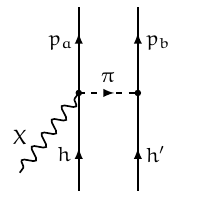}\label{fig:diagrams_2p2h_mec:sea1}}
  \subfloat[]{\includegraphics[width=0.19\textwidth]{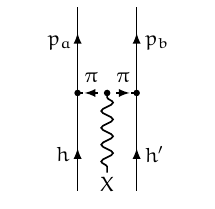}\label{fig:diagrams_2p2h_mec:pif}}
  \subfloat[]{\includegraphics[width=0.19\textwidth]{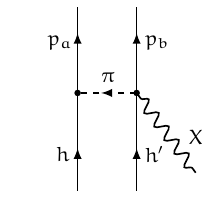}\label{fig:diagrams_2p2h_mec:sea2}}
  \\
  \subfloat[]{\includegraphics[width=0.19\textwidth]{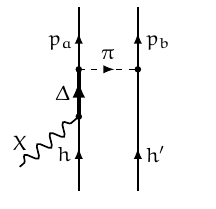}\label{fig:diagrams_2p2h_mec:delta1}}
  \subfloat[]{\includegraphics[width=0.19\textwidth]{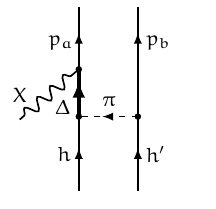}\label{fig:diagrams_2p2h_mec:delta2}}
  \subfloat[]{\includegraphics[width=0.19\textwidth]{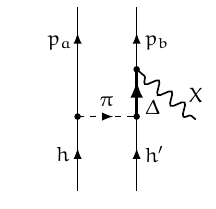}\label{fig:diagrams_2p2h_mec:delta3}}
  \subfloat[]{\includegraphics[width=0.19\textwidth]{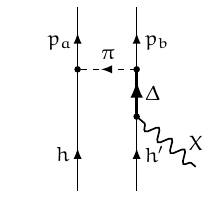}\label{fig:diagrams_2p2h_mec:delta4}}
  \caption{Graphical representation of the diagrams involved in the two-nucleon knock-out processes induced with the meson-exchange currents dynamics: (a)(c) seagull, (b) pion-in-flight, (d)(g) direct (s-channel) $\Delta$-mediated currents, and (e)(f) crossed (u-channel) $\Delta$-mediated currents. Labels $h,h'$ indicate occupied single-particle states, and $p_a,p_b$ denote unbound single-particle states in the mean-field potential.}
  \label{fig:diagrams_2p2h_mec}
\end{figure*}

We consider two main MEC types: those that involve nucleon resonances and those that do not. At lower energy transfers, the dominant contribution arises from the latter ones, commonly referred to as the \textit{seagull} and \textit{pion-in-flight} currents. These two currents differ based on the coupling point of the boson---either at the $\pi NN$ vertex or directly to the exchanged pion, as illustrated in the top row of Fig.~\ref{fig:diagrams_2p2h_mec}. Notably, an exchanged pion introduces a parity change compared to the one-body currents defined in Eqs.~\eqref{eq:ia_time} and~\eqref{eq:ia_space}, reflecting their distinct role in the electromagnetic response. In general, a non-relativistic reduction of the pion-production amplitude yields terms of $\mathcal{O}(1)$ for the space part and $\mathcal{O}(\sfrac{|\mathbf{q}|}{M_N})$ for the time part of the vector current~\cite{Towner:1987zz}. We retain the dominant, space part of the electromagnetic meson-exchange currents.

The elementary building block for constructing meson-exchange currents is the $\pi NN$ interaction vertex, for which we adopt a pseudovector coupling derived from an effective Lagrangian in the low-energy limit~\cite{Ericson:1988gk}:
\begin{equation}
    \label{eq:lagrangian_piNN}
    \mathcal{L}_{\pi N N, i}(\mathbf{r}) =  \frac{f_{\pi N N}}{m_{\pi}} \left(\bm{\sigma}_i \cdot \bm{\nabla} \right)(\bm{\tau}_i \cdot \bm{\pi}(\mathbf{r})) \delta^{(3)}(\mathbf{r}_i-\mathbf{r}) ,
\end{equation}
where $\bm{\pi}(\mathbf{r})$ denotes the isovector pion field. The coupling constant is set via $f_{\pi NN}^2 / (4\pi) = 0.08$. By gauging the $\pi NN$ Lagrangian through minimal substitution, we generate the seagull current in the form
\begin{widetext}
\begin{equation}
\begin{split}
\label{eq:mec_bare_sea}
    \hat{\bm{J}}_{ij}^{[2], \mathrm{sea}} (\mathbf{r}) =
    - \left(\frac{f_{\pi N N}}{m_{\pi}}\right)^{2}
    (\mathrm{I}_{V,ij})_{z}
    \delta^{(3)}(\mathbf{r}_i-\mathbf{r}) \bm{\sigma}_{i} \left(\bm{\sigma}_{j} \cdot \bm{\nabla}_{j}\right)
    \frac{e^{-m_\pi |\mathbf{r}_{j} - \mathbf{r}|}}{4 \pi |\mathbf{r}_{j} - \mathbf{r}|} + \left(i \leftrightarrow j\right) ,
\end{split}
\end{equation}
where the two-body isovector operator is defined as $\mathbf{I}_{V,ij} = \bm{\tau}_i \times \bm{\tau}_j$. Then, we construct the pion-in-flight current by gauging the free pion Lagrangian through minimal substitution and obtain
\begin{equation}
\begin{split}
\label{eq:mec_bare_pif}
    \hat{\bm{J}}_{ij}^{[2], \mathrm{pif}} (\mathbf{r}) =
    - \left(\frac{f_{\pi N N}}{m_{\pi}}\right)^{2}
    (\mathrm{I}_{V,ij})_{z}
    \left(\bm{\sigma}_{i} \cdot \bm{\nabla}_{i}\right) \left(\bm{\sigma}_{j} \cdot \bm{\nabla}_{j}\right)
    \left( \bm{\nabla}_{i} - \bm{\nabla}_{j} \right)
    \frac{e^{-m_\pi |\mathbf{r}_{i} - \mathbf{r}|}}{4 \pi |\mathbf{r}_{i} - \mathbf{r}|}
    \frac{e^{-m_\pi |\mathbf{r}_{j} - \mathbf{r}|}}{4 \pi |\mathbf{r}_{j} - \mathbf{r}|} .
\end{split}
\end{equation}
These two currents lead to the well-known one-pion exchange potential
\begin{equation}
\label{eq:opep}
    \hat{V}_{\pi,ij} (\mathbf{r}_{ij}) =
    \frac{1}{3} \left(\frac{f_{\pi N N}}{m_{\pi}}\right)^{2}
    \left( \bm{\tau}_i \cdot \bm{\tau}_j \right)
    \left\{ \bm{\sigma}_i \cdot \bm{\sigma}_j + \hat{S}_{ij}(\mathbf{r}_{ij})
    \left[1 + \frac{3}{m_\pi |\mathbf{r}_{ij}|}
    + \frac{3}{(m_\pi |\mathbf{r}_{ij}|)^2} \right] \right\}
    \frac{e^{-m_\pi |\mathbf{r}_{ij}|}}{|\mathbf{r}_{ij}|},
\end{equation}
\end{widetext}
where the tensor operator is defined as
\begin{equation}
\label{eq:tensor_operator}
    \hat{S}_{ij}(\mathbf{r}_{ij}) = \frac{3}{|\mathbf{r}_{ij}|^2} (\bm{\sigma}_i\cdot\mathbf{r}_{ij}) (\bm{\sigma}_j\cdot\mathbf{r}_{ij}) - (\bm{\sigma}_i\cdot\bm{\sigma}_j) .
\end{equation}
Finally, per analogy with the one-body currents of Eqs.~\eqref{eq:ia_time} and~\eqref{eq:ia_space}, we dress the photon isovector coupling, motivated by the vector-meson dominance model, with the same $F_1^V(Q^2)$ form factor in both the seagull and pion-in-flight operators.

The meson-exchange currents constructed in this manner, together with the above-mentioned one-body currents, satisfy the conserved vector current (CVC) condition at the Born level~\cite{Towner:1987zz}. This requirement also constrains any additional transverse contributions to satisfy CVC, i.e., $\bm{\nabla} \cdot \bm{J}^{[2]}_{ij} (\mathbf{r}) = 0$, on their own. As a consequence, the associated two-body currents are determined by the underlying pion-exchange interaction and are therefore often referred to in the literature as \textit{model-independent} pion-exchange current operators~\cite{Mathiot:1989vw}, since they do not introduce any free parameters beyond those specified in the interaction of Eq.~\eqref{eq:opep}.

In contrast to many relativistic formulations of meson-exchange currents~\cite{Amaro:2002mj}, we do not explicitly include the so-called correlation (or nucleon-pole) currents in the diagrams of Fig.~\ref{fig:diagrams_2p2h_mec}. As argued in Ref.~\cite{Towner:1987zz}, the non-relativistic reduction of the nucleon propagator in these diagrams involves several subtleties. Firstly, the positive-frequency part does not yield a true two-body operator, but rather a product of a one-body current and the nucleon-nucleon potential. Those contributions are naturally incorporated into our impulse approximation through the use of HF wavefunctions. Secondly, the negative-frequency part of the propagator in a pseudoscalar coupling framework is mathematically equivalent to the seagull current derived from minimal substitution within our pseudovector coupling. Consequently, by adopting the effective Lagrangian of Eq.~\eqref{eq:lagrangian_piNN}, we consistently account for these effects while avoiding double counting of correlation currents.

A further refinement of the MECs is required at high momentum transfers, where the assumption of point-like nucleons breaks down. Then, hadronic substructure becomes relevant. We regularize the $\pi NN$ vertices by introducing a monopole hadronic form factor:
\begin{equation}
\label{eq:monoform}
    \Gamma_{\pi}\left(|\mathbf{q}|^{2}\right)=\frac{\Lambda_{\pi}^{2}-m_{\pi}^{2}}{|\mathbf{q}|^{2}+\Lambda_{\pi}^{2}},
\end{equation}
with a cut-off mass $\Lambda_\pi = 1250 \ \mathrm{MeV}/c^2$. This monopole form has a natural physical interpretation as introducing a \textit{heavy pion} mediating the $\pi N$ coupling.  While this regularization is essential for controlling short-distance behavior and ensuring numerical stability, it formally violates the CVC condition once applied to the seagull and pion-in-flight operators.

A commonly adopted prescription to restore CVC is replacing the product of form factors with~\cite{Mathiot:1984bs,Towner:1987zz}:
\begin{equation}
\begin{split}
    \mathcal{F}(|\mathbf{q}_{i}|^{2},|\mathbf{q}_{j}|^{2}) & =
    \Gamma_{\pi}(|\mathbf{q}_{i}|^{2})\Gamma_{\pi}(|\mathbf{q}_{j}|^{2}) \\
    & \times \bigg( 1 +
    \frac{|\mathbf{q}_{i}|^{2} + m_{\pi}^{2}}{|\mathbf{q}_{j}|^{2} + \Lambda_{\pi}^{2}} + \frac{|\mathbf{q}_{j}|^{2} + m_{\pi}^{2}}{|\mathbf{q}_{i}|^{2} + \Lambda_{\pi}^{2}} \bigg).
\end{split}
\end{equation}
Beyond the two vertex form factors, the additional terms in $\mathcal{F}$ represent the boson coupling directly to the exchanged heavy pion. In practice, we find that the additional terms reintroduce substantial high-momentum contributions, leading to numerical instabilities. Since numerical convergence of the matrix elements is crucial for meaningful comparison with experimental data, we apply the monopole form factors $\Gamma_\pi$ of Eq.~\eqref{eq:monoform} at the vertices. This choice violates the continuity equation at large $Q^2$, but ensures stable and well-behaved predictions for the two-body current contributions.

The resulting position-space expression for the seagull and pion-in-flight contributions combined into a single non-resonant meson-exchange current operator reads~\cite{Ryckebusch:1997gn}
\begin{widetext}
\begin{equation}
\begin{split}
\label{eq:mec_sea_pif}
    \hat{\bm{J}}_{ij}^{[2], \mathrm{sea}+\mathrm{pif}} (\mathbf{r}) = & 
    - \left(\frac{f_{\pi N N}}{m_{\pi}}\right)^{2}
    F_{1}^{V}(Q^2) (\mathrm{I}_{V,ij})_{z} \\
    & \times
    \begin{aligned}
        & \bigg\{ \Gamma_{\pi}(|\mathbf{q}_{i}|^{2}) \delta^{(3)}(\mathbf{r}_i-\mathbf{r}) \bm{\sigma}_{i} \left(\bm{\sigma}_{j} \cdot \bm{\nabla}_{j}\right)
        \frac{e^{-m_\pi |\mathbf{r}_{j} - \mathbf{r}|}}{4 \pi |\mathbf{r}_{j} - \mathbf{r}|}
        - \Gamma_{\pi}(|\mathbf{q}_{j}|^{2}) \delta^{(3)}(\mathbf{r}_j-\mathbf{r}) \bm{\sigma}_{j} \left(\bm{\sigma}_{i} \cdot \bm{\nabla}_{i}\right)
        \frac{e^{-m_\pi |\mathbf{r}_{i} - \mathbf{r}|}}{4 \pi |\mathbf{r}_{i} - \mathbf{r}|} \\
        & + \Gamma_{\pi}(|\mathbf{q}_{i}|^{2}) \Gamma_{\pi}(|\mathbf{q}_{j}|^{2}) \left(\bm{\sigma}_{i} \cdot \bm{\nabla}_{i}\right) \left(\bm{\sigma}_{j} \cdot \bm{\nabla}_{j}\right)
        \left( \bm{\nabla}_{i} - \bm{\nabla}_{j} \right)
        \frac{e^{-m_\pi |\mathbf{r}_{i} - \mathbf{r}|}}{4 \pi |\mathbf{r}_{i} - \mathbf{r}|}
        \frac{e^{-m_\pi |\mathbf{r}_{j} - \mathbf{r}|}}{4 \pi |\mathbf{r}_{j} - \mathbf{r}|} \bigg\} .
    \end{aligned}
\end{split}
\end{equation}
\end{widetext}

\subsubsection{Isobar degrees of freedom}

A significant photon-nucleus interaction contribution arises when we include the ability of the photon to excite the target nucleon into a resonant state. Focusing on the prominent $\Delta(1232)$ isobar resonance ($J^\pi = \sfrac{3}{2}^+$, $I = \sfrac{3}{2}$), we construct the corresponding meson-exchange currents as illustrated in the diagrams of Figs.~\ref{fig:diagrams_2p2h_mec:delta1}-\ref{fig:diagrams_2p2h_mec:delta4}. Commonly referred to as $\Delta$-currents, they provide a substantial strength in scattering experiments and are essential for a reliable description of two-nucleon knock-out processes~\cite{Ryckebusch:1997gn}. These currents are constructed by coupling the probe to a $\Delta$-production vertex, followed by a pion exchange with a spectator nucleon. This process is mediated by the isobar through both direct and crossed pole diagrams, which we evaluate using resonant and non-resonant propagators, respectively.

Unlike the seagull and pion-in-flight currents, which are constrained by the continuity 
equation and the one-pion exchange potential, the non-relativistic $\Delta$-isobar currents are purely transverse~\cite{Towner:1987zz}. Consequently, they are often characterized as \textit{model-dependent}~\cite{Mathiot:1989vw} and require the introduction of the $N(\sfrac{1}{2})\to\Delta(\sfrac{3}{2})$ transition spin ($\mathbf{S}$) and isospin ($\mathbf{T}$) operators. Thus, we encounter two new interaction vertices: $\pi N \Delta$ and $\gamma N \Delta$. The former describes a pion-induced nucleon excitation into a $\Delta$ resonance, while the latter describes the dominant magnetic M1 channel of the electromagnetic excitation of the nucleon target. To evaluate these contributions, we introduce effective Lagrangians of the form~\cite{Ericson:1988gk}
\begin{align}
\begin{split}
    \mathcal{L}_{\pi N \Delta, i} (\mathbf{r}) = & \frac{f_{\pi N \Delta}}{m_{\pi}} \left( \mathbf{S}^{\dagger}_i \cdot \bm{\nabla} \right) \left( \mathbf{T}^{\dagger}_i \cdot \bm{\pi}(\mathbf{r}) \right) \\
    & \times \delta^{(3)}(\mathbf{r}_i - \mathbf{r}) + \mathrm{h.c.} ,
\end{split}\\
\begin{split}
    \mathcal{L}_{\gamma N \Delta, i} (\mathbf{r}) = & \frac{f_{\gamma N \Delta}}{m_{\pi}}G_{\gamma N \Delta}(Q^{2}) \left( \bm{\nabla} \times \mathbf{S}^{\dagger}_i \right) \cdot \mathbf{A}(\mathbf{r}) \\
    & \times (\mathrm{T}^{\dagger}_{i})_z \delta^{(3)}(\mathbf{r}_i - \mathbf{r}) + \mathrm{h.c.} ,
\end{split}
\end{align}
where $\mathbf{A}(\mathbf{r})$ is the external electromagnetic field, and the coupling constants are $f^2_{\pi N \Delta} / (4\pi) = 0.37$ and $f_{\gamma N \Delta} = 0.12$~\cite{Ryckebusch:1997gn}. The electromagnetic form factor $G_{\gamma N \Delta} (Q^2)$ is parametrized as~\cite{Marcucci:1998tb}
\begin{equation}
    G_{\gamma N \Delta}(Q^{2})=\frac{1}{\left(1+\nicefrac{Q^{2}}{\Lambda_{1}^{2}}\right)} \frac{1}{\left(1+\nicefrac{Q^{2}}{\Lambda_{2}^{2}}\right)},
\end{equation}
where $\Lambda_1 = 840 \ \mathrm{MeV}/c^2$ and $\Lambda_2 = 1200 \ \mathrm{MeV}/c^2$. Note that this form factor decreases with $Q^2$ faster than the nucleon dipole form~\cite{Stoler:1993yk,Frolov:1998pw}.

To derive the two-body current operator from these vertices, one defines an isobar propagator that mediates the excitation. In a non-relativistic framework, while the \textit{static approximation} provides a natural starting point, it lacks the characteristic resonant behavior observed in experimental cross sections. Therefore, we adopt a prescription that accounts for the finite lifetime of the $\Delta(1232)$ and its in-medium properties, reflecting the fact that the $\Delta$ is a spatially more extended object than a nucleon.

\begin{figure}[ht!]
    \centering
    \subfloat[Direct term ($s$-channel)]{\includegraphics[width=0.19\textwidth]{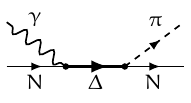}\label{fig:delta_resonance:direct}}
    \subfloat[Crossed term ($u$-channel)]{\includegraphics[width=0.19\textwidth]{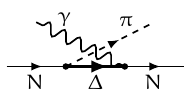}\label{fig:delta_resonance:crossed}}
    \caption{Graphical representation of the $\Delta$-resonance excitation with an external photon, and its de-excitation.}
    \label{fig:delta_resonance}
\end{figure}

Defining field properties for particles with $J > \sfrac{1}{2}$ and non-vanishing mass is challenging. Non-uniqueness and gauge freedom lead to an inevitable complexity, usually expressed in the Rarita-Schwinger formalism~\cite{Rarita:1941mf}. The two $\Delta$ propagators appearing in the diagrams of Fig.~\ref{fig:diagrams_2p2h_mec} correspond to the direct and crossed $\Delta$-pole terms, shown schematically in Fig.~\ref{fig:delta_resonance}. To describe the isobar dynamics, it is typical to adopt the following propagators~\cite{Ryckebusch:1999gu}
\begin{align}
    G_\Delta^\mathrm{res} & = \frac{1}{M_\Delta - \sqrt{s} - \sfrac{i}{2}\Gamma_\Delta^\mathrm{res} + V_\Delta}, \label{eq:delta:propagators:res} \\
    G_\Delta^\mathrm{nres} & = \frac{1}{M_\Delta - \sqrt{u}}, \label{eq:delta:propagators:nres}
\end{align}
where $\sqrt{s}$ and $\sqrt{u}$ are the invariant energies available in the resonant and non-resonant channels, respectively. The resonant propagator includes the energy-dependent free width $\Gamma_\Delta^\mathrm{res}$, which we parametrize as~\cite{Vanderhaeghen:1995fe,Dekker:1994yc}
\begin{equation}
    \Gamma_\Delta^\mathrm{res} = \frac{1}{3} \frac{f_{\pi N \Delta}^2}{4\pi} \frac{|\mathbf{p}_\pi|^3}{m_\pi^2} \frac{M_N + E_\pi}{\sqrt{s}} ,
\end{equation}
expressed in center-of-momentum frame variables. Additionally, we account for in-medium effects by introducing the potential $V_\Delta$, which shifts the resonance position and modifies the width to reflect the nuclear environment. For additional details on the $\Delta$ decay width and unitarization within the Ghent framework, see Ref.~\cite{Hooft:2026yyg}.

We obtain the current by performing a summation over the intermediate $\Delta$ states, which effectively reduces the transition operators $\mathbf{S}$ and $\mathbf{T}$ to standard Pauli matrices in the nucleon subspace. This is achieved through the relation:
\begin{equation}
    S_a S^\dagger_b = \frac{2}{3} \delta_{ab} - \frac{i}{3} \epsilon_{abc} \sigma_c
\end{equation}
and an analogous expression for the isospin operators $T_a T^\dagger_b$. By grouping the terms into resonant and non-resonant combinations $(G_\Delta^\mathrm{res} \pm G_\Delta^\mathrm{nres})$, we isolate the contributions to the symmetric and antisymmetric components of the current. Finally, we regularize the $\pi NN$ and $\pi N \Delta$ vertices with the monopole form factor $\Gamma_\pi(|\mathbf{q}|^2)$ of Eq.~\eqref{eq:monoform} and obtain~\cite{Ryckebusch:1997gn}
\begin{widetext}
\begin{equation}
\begin{split}
\label{eq:mec_delta}
    \hat{{J}}^{[2],\Delta}_{ij}(\mathbf{r}) = & \ \frac{1}{9} \frac{f_{\pi NN} f_{\pi N \Delta} f_{\gamma N \Delta}}{m_\pi^3} G_{\gamma N \Delta}(Q^2) \\
    & \times \Bigg\{ (G_\Delta^\mathrm{res} + G_\Delta^\mathrm{nres})
        \Bigg( \left[ (\mathrm{I}_{V,ij})_{z} \Gamma_\pi^2 (|\mathbf{q}_i|^2)
        \left( \bm{\sigma}_{i} \cdot \bm{\nabla}_{i} \right)
        \left( \bm{\nabla} \times \left(\bm{\sigma}_{j} \times \bm{\nabla}_{i}\right) \right)
        \frac{e^{-m_\pi |\mathbf{r}_i - \mathbf{r}|}}{4\pi |\mathbf{r}_i - \mathbf{r}|}
        \delta^{(3)} ( \mathbf{r}_j - \mathbf{r} )
        \right. \\
    & \phantom{\times \Bigg\{ (G_\Delta^\mathrm{res} + G_\Delta^\mathrm{nres}) \Bigg(}
        \left. - (\mathrm{I}_{V,ij})_{z} \Gamma_\pi^2 (|\mathbf{q}_j|^2)
        \left( \bm{\sigma}_{j} \cdot \bm{\nabla}_{j} \right)
        \left( \bm{\nabla} \times \left( \bm{\sigma}_{i} \times \bm{\nabla}_{j}\right) \right)
        \delta^{(3)} ( \mathbf{r}_i - \mathbf{r} )
        \frac{e^{-m_\pi |\mathbf{r}_j - \mathbf{r}|}}{4\pi |\mathbf{r}_j - \mathbf{r}|}
        \right] \\
    & \phantom{\times \Bigg\{ (G_\Delta^\mathrm{res} + G_\Delta^\mathrm{nres})}
        + \left[ 4 (\tau_i)_z \Gamma_\pi^2 (|\mathbf{q}_i|^2)
        \left( \bm{\sigma}_{i} \cdot \bm{\nabla}_{i} \right)
        \left( \bm{\nabla} \times \bm{\nabla}_i \right)
        \frac{e^{-m_\pi |\mathbf{r}_i - \mathbf{r}|}}{4\pi |\mathbf{r}_i - \mathbf{r}|}
        \delta^{(3)} ( \mathbf{r}_j - \mathbf{r} )
        \right. \\
    & \phantom{\times \Bigg\{ (G_\Delta^\mathrm{res} + G_\Delta^\mathrm{nres}) \Bigg(}
        + \left. 4 (\tau_j)_z \Gamma_\pi^2 (|\mathbf{q}_j|^2)
        \left( \bm{\sigma}_{j} \cdot \bm{\nabla}_{j} \right)
        \left( \bm{\nabla} \times \bm{\nabla}_j \right)
        \delta^{(3)} ( \mathbf{r}_i - \mathbf{r} )
        \frac{e^{-m_\pi |\mathbf{r}_j - \mathbf{r}|}}{4\pi |\mathbf{r}_j - \mathbf{r}|}
        \right] \Bigg) \\
    & \phantom{\times} + (G_\Delta^\mathrm{res} - G_\Delta^\mathrm{nres})
        \Bigg( \left[ 2i (\mathrm{I}_{V,ij})_{z} \Gamma_\pi^2 (|\mathbf{q}_i|^2)
        \left( \bm{\sigma}_{i} \cdot \bm{\nabla}_{i} \right)
        \left( \bm{\nabla} \times \bm{\nabla}_i \right)
        \frac{e^{-m_\pi |\mathbf{r}_i - \mathbf{r}|}}{4\pi |\mathbf{r}_i - \mathbf{r}|}
        \delta^{(3)} ( \mathbf{r}_j - \mathbf{r} )
        \right. \\
    & \phantom{\times + (G_\Delta^\mathrm{res} - G_\Delta^\mathrm{nres}) \Bigg(}
        \left. - 2i (\mathrm{I}_{V,ij})_{z} \Gamma_\pi^2 (|\mathbf{q}_j|^2)
        \left( \bm{\sigma}_{j} \cdot \bm{\nabla}_{j} \right)
        \left( \bm{\nabla} \times \bm{\nabla}_j \right)
        \delta^{(3)} ( \mathbf{r}_i - \mathbf{r} )
        \frac{e^{-m_\pi |\mathbf{r}_j - \mathbf{r}|}}{4\pi |\mathbf{r}_j - \mathbf{r}|}
        \right] \\
    & \phantom{\times + (G_\Delta^\mathrm{res} - G_\Delta^\mathrm{nres})}
        - \left[ 2i (\tau_i)_z \Gamma_\pi^2 (|\mathbf{q}_i|^2)
        \left( \bm{\sigma}_{i} \cdot \bm{\nabla}_{i} \right)
        \left( \bm{\nabla} \times \left(\bm{\sigma}_{j} \times \bm{\nabla}_{i}\right) \right)
        \frac{e^{-m_\pi |\mathbf{r}_i - \mathbf{r}|}}{4\pi |\mathbf{r}_i - \mathbf{r}|}
        \delta^{(3)} ( \mathbf{r}_j - \mathbf{r} )
        \right. \\
    & \phantom{\times + (G_\Delta^\mathrm{res} - G_\Delta^\mathrm{nres}) \Bigg(}
        + \left. 2i (\tau_j)_z \Gamma_\pi^2 (|\mathbf{q}_j|^2)
        \left( \bm{\sigma}_{j} \cdot \bm{\nabla}_{j} \right)
        \left( \bm{\nabla} \times \left( \bm{\sigma}_{i} \times \bm{\nabla}_{j}\right) \right)
        \delta^{(3)} ( \mathbf{r}_i - \mathbf{r} )
        \frac{e^{-m_\pi |\mathbf{r}_j - \mathbf{r}|}}{4\pi |\mathbf{r}_j - \mathbf{r}|}
        \right] \Bigg) \Bigg\}
\end{split}
\end{equation}
\end{widetext}

\begin{figure}[t]
  \centering
  \includegraphics[width=0.48\textwidth]{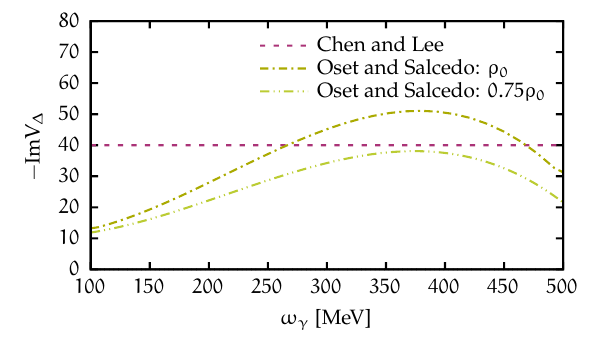}
  \caption{Parametrizations of the imaginary part of the $\Delta$-resonance nuclear potential $V_\Delta$. The purple dashed line presents the constant parametrization of Ref.~\cite{Chen:1988xq}, while the olive dash-dot and pear dash-dot-dot lines provide the density-dependent parametrization of Ref.~\cite{Oset:1987re} at the saturation density and $0.75$ of that value, respectively.}
  \label{fig:delta_inmedium}
\end{figure}

Among all pion-related two-body current contributions, the $\Delta$-current is the only one explicitly shown to exhibit strong medium dependence~\cite{Ryckebusch:1999gu}. Typical values for the $\Delta$ self-energy in the medium ($\Sigma_\Delta$) are $V_\Delta^\mathrm{Chen\&Lee} = -30 -i40 \ [\mathrm{MeV}]$~\cite{Chen:1988xq}, as utilized in the calculations of Ref~\cite{Ryckebusch:1997gn}. E.~Oset and L.~Salcedo calculated this quantity for pion photo-production in a Fermi gas as a function of photon energy~\cite{Oset:1987re}. To adapt this to electron scattering, we introduce the equivalent real-photon energy
\begin{equation}
    \omega_\gamma = \omega + \frac{q^2}{2M_N} ,
\end{equation}
which approximates the energy transfer required to excite the resonance in the presence of a non-zero three-momentum transfer $|\mathbf{q}|$. This potential is sensitive to the local nuclear density relative to the saturation value $(\rho/\rho_0)$. For finite nuclei, it has been argued that an effective density of $\rho=0.75\rho_0$ provides a more accurate representation of the nuclear environment~\cite{Nieves:1993ev}. However, in the present work, we follow the arguments of J.~Ryckebusch \textit{et al.}~\cite{Ryckebusch:1999gu} and take $\rho = \rho_0$. Fig.~\ref{fig:delta_inmedium} illustrates the resulting imaginary part of the $\Delta$ potential as a function of the effective energy transfer. The direct broadening of the $\Delta$ resonance width in nuclei has been extensively studied, for instance by M.~Hjorth-Jensen \textit{et al.}~\cite{Hjorth-Jensen:1993qiu}.

Calculating the propagators $G_\Delta^\mathrm{res}$ and $G_\Delta^\mathrm{nres}$ requires the Mandelstam variables $s$ and $u$. To define the kinematics of the excitation processes, we consider the direct term illustrated in Fig.~\ref{fig:delta_resonance:direct}, the external boson transfers energy ($\omega$) and momentum ($|\mathbf{q}|$) directly to the nucleon. Assuming the target nucleon, with hole (removal) energy $\epsilon_h$, is initially at rest, we obtain the invariant mass squared:
\begin{equation}
    s = q^2 + (M_N - \epsilon_h)^2 + 2\omega(M_N - \epsilon_h) .
\label{eq:delta:mandelstam_s}
\end{equation}
To maintain consistency with pion photo-production models, one could introduce an effective quantity~\cite{Wilbois:1996zz}
\begin{equation}
    s_\gamma = (M_N - \epsilon_h)^2 + 2\omega_\gamma(M_N - \epsilon_h)
\label{eq:delta:mandelstam_s2}
\end{equation}
to translate the process to a real-photon equivalent. However, in the kinematic conditions of the present study, the correction is close to negligible ($s-s_\gamma \approx 0.015q^2$) and the prescription of Eq.~\eqref{eq:delta:mandelstam_s}~\cite{Ryckebusch:1999gu} can be utilized.

The evaluation of the $u = (q - p_{N_a})^2$ variable is more complex, as the kinematic constraints of inclusive $A(e,e')$ do not provide direct access to the dynamics of the intermediate isobar. While Ref.~\cite{Ryckebusch:1999gu} assumes the $\Delta$ is at rest and receives the full momentum transferred to the system ($\vec{p}_{N_a} \simeq \vec{q}$), this approach is prone to unphysical divergences as $\sqrt{u} \to M_\Delta$. To resolve this, we introduce more dynamical information into the formalism by considering the momentum conservation of the final-state nucleons:
\begin{equation}
    \mathbf{P}_{hh'} + \mathbf{q} = \mathbf{p}_{N_a} + \mathbf{p}_{N_b},
\end{equation}
where $\mathbf{P}_{hh'}$ is the momentum of the initial $NN$ center-of-mass system. We adopt quasi-deuteron kinematics~\cite{Gottfried:1958} by assuming back-to-back initial nucleons ($|\mathbf{P}_{hh'}| \approx 0$). By averaging over the final nucleon momenta ($\mathbf{p}_{N_a} \simeq \mathbf{q}/2$), we obtain a better-constrained estimate for $u$:
\begin{equation}
    u = \left( \sqrt{M_N^2 + |\mathbf{q}|^2/4} - \omega \right)^2 - \left( \frac{|\mathbf{q}|^2}{4} \right) .
\label{eq:delta:mandelstam_u}
\end{equation}
This prescription provides a significantly more stable treatment of the non-resonant background than the static approach used in Ref.~\cite{Ryckebusch:1999gu}.

To further improve the treatment of relativistic effects in the $\Delta$-excitation process, it is necessary to account for the pole structure in the direct and crossed channels, which take the form of $(M_\Delta^2 - s)^{-1}$ and $(M_\Delta^2 - u)^{-1}$, respectively~\cite{Ericson:1988gk}. Following the prescription of J.~Dekker \textit{et al.}~\cite{Dekker:1994yc}, we utilize the propagators that maintain the relativistic Breit-Wigner distribution:
\begin{align}
    G_\Delta^\mathrm{res} & = \frac{2M_\Delta}{M_\Delta^2 - s - iM_\Delta\Gamma_\Delta^\mathrm{res} + 2 M_\Delta V_\Delta}, \\
    G_\Delta^\mathrm{nres} & = \frac{2M_\Delta}{M_\Delta^2 - u}.
\end{align}
Although these corrections yield relatively minor changes in the $s$-channel compared to a linear expansion, the overall improvements in the $u$-channel, particularly when combined with the kinematic prescription of Eq.~\eqref{eq:delta:mandelstam_u}, provide a more stable and physically motivated description of the isobar contribution~\cite{Niewczas:2023phd}.

\begin{figure*}[t]
  \centering
  \includegraphics[width=1.0\textwidth]{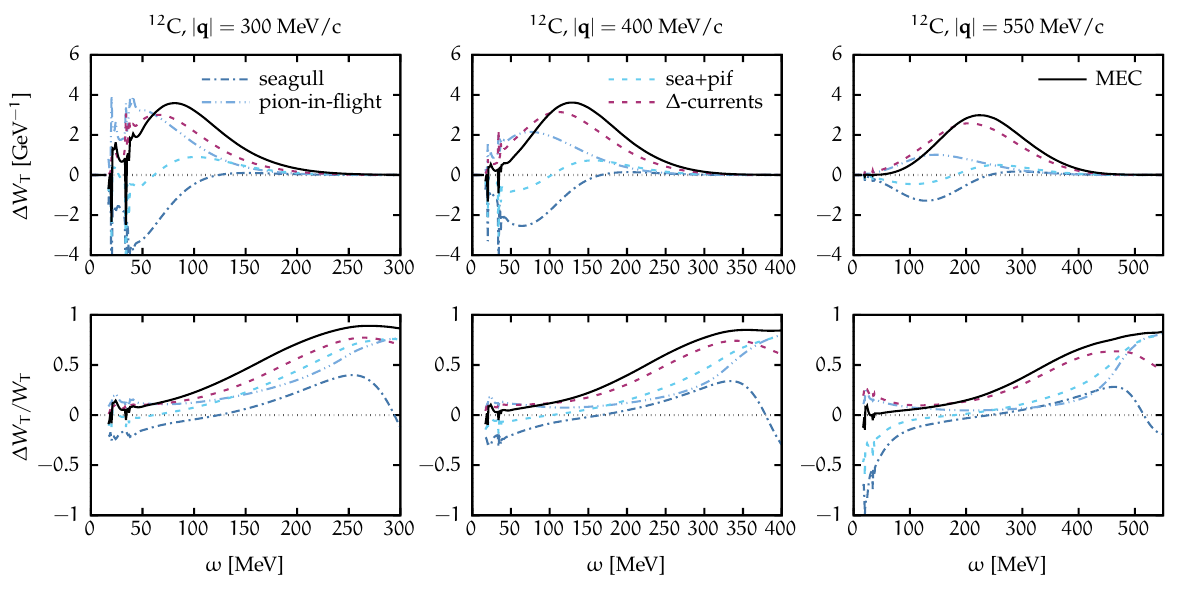}
  \caption{Energy dependence of the absolute and fractional MEC corrections to the 1p1h hadronic $^{12}\mathrm{C}(e,e')$ responses for fixed three-momentum transfer. The blue dash-dot, light blue dash-dot-dot, and cyan dashed lines present the seagull, pion-in-flight, and combined contributions, respectively. The purple dashed line provides the effect of $\Delta$-currents, while the solid black---the total meson-exchange currents contribution, both using the $\Delta$-resonance self-energy parametrization of Ref.~\cite{Chen:1988xq}.}
  \label{fig:incl_q_corr2_mecs}
\end{figure*}

We also wish to acknowledge the evolution of the approximations within our framework. In earlier studies, such as Ref.~\cite{Ryckebusch:1993tf}, the propagators were treated in a \textit{static} limit, defined as:
\begin{align}
    G_\Delta^\mathrm{res} & = \frac{1}{M_\Delta - M_N}, \\
    G_\Delta^\mathrm{nres} & = 0 .
\end{align}
This approach significantly restricted the number of contributing spin and isospin structures in Eq.~\eqref{eq:mec_delta}. A subsequent, \textit{resonating} approach, proposed by J.~Dekker \textit{et al.}~\cite{Dekker:1994yc} and extended in Ref.~\cite{VanderSluys:1995rp}, introduced a more dynamic prescription:
\begin{align}
    \begin{split}
        G_\Delta^\mathrm{res} + G_\Delta^\mathrm{nres} & = \frac{1}{M_\Delta - M_N - \omega - \sfrac{i}{2}\Gamma_\Delta^\mathrm{res}} \\
        & + \frac{1}{M_\Delta - M_N + \omega},
    \end{split} \\
    G_\Delta^\mathrm{res} - G_\Delta^\mathrm{nres} & = 0 .
\end{align}
While this accounted for the resonance energy, it implied operator structures of the static case by assuming symmetry between the direct and crossed channels. The calculation, utilizing the full range of terms enabled by the propagators in Eqs.~\eqref{eq:delta:propagators:res} and~\eqref{eq:delta:propagators:nres}, was presented in Ref.~\cite{Ryckebusch:1997gn}.

Finally, we address the treatment of the $\Delta$-isobar width in inclusive reactions. While unified microscopic frameworks can derive both two-body meson-exchange currents and single-pion production (SPP) from a common in-medium $\Delta$ self-energy~\cite{Gil:1997bm}, such consistency is rarely enforced in practice. Modeling these two sectors independently risks double-counting cross section strength in the $\Delta$-resonance region. To avoid such double-counting, it has been argued that the imaginary part of the propagator should be omitted in the MEC sector~\cite{Megias:2018ujz,RuizSimo:2016rtu}, following the relativistic two-body current framework developed in Ref.~\cite{DePace:2003spn}. In this view, the imaginary part physically represents the $\Delta$ decay to an on-shell pion, a channel that is typically accounted for by separate inelastic or pion-production modeling. The self-energy $\Sigma_\Delta$ is inherently channel-blind, shaping both $\Re(G_\Delta)$ and $\Im(G_\Delta)$. Discarding the latter breaks the relation between the real and imaginary parts required by unitarity and removes genuine resonant strength from the 2p2h channel itself. Moreover, since $[\Re(G_\Delta)]^2$ vanishes at $\sqrt{s} \approx M_\Delta+V_\Delta$, this prescription forces the cross section to zero exactly at resonance. This could cause the dip-region strength to appear dominant over the (artificially suppressed) resonance peak, rather than reflecting a genuine physical effect. We therefore retain the full complex propagator in our calculations.

\subsubsection{Hadronic responses}

\begin{figure*}[ht!]
  \centering
  \includegraphics[width=1.0\textwidth]{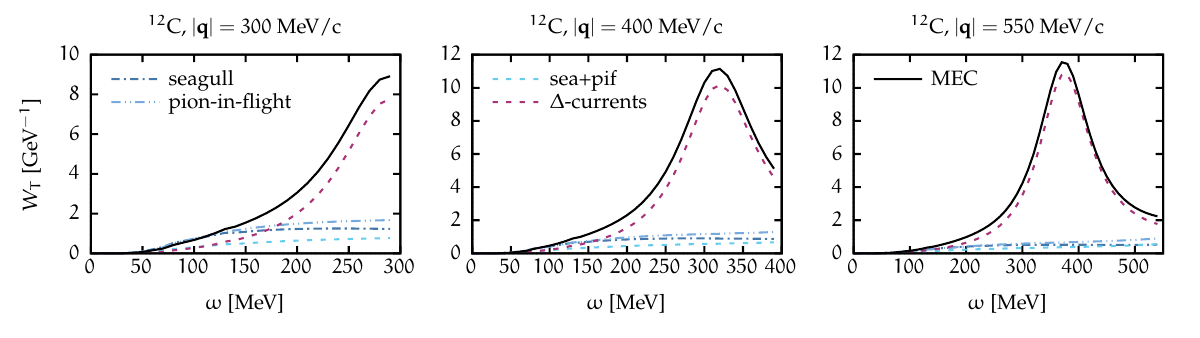}
  \caption{Energy dependence of 2p2h hadronic $^{12}\mathrm{C}(e,e')$ responses induced by meson-exchange currents for fixed three-momentum transfer. Line conventions as in Fig.~\ref{fig:incl_q_corr2_mecs}.}
  \label{fig:incl_q_2p2h_mecs}
\end{figure*}

\begin{figure*}[ht!]
  \centering
  \includegraphics[width=1.0\textwidth]{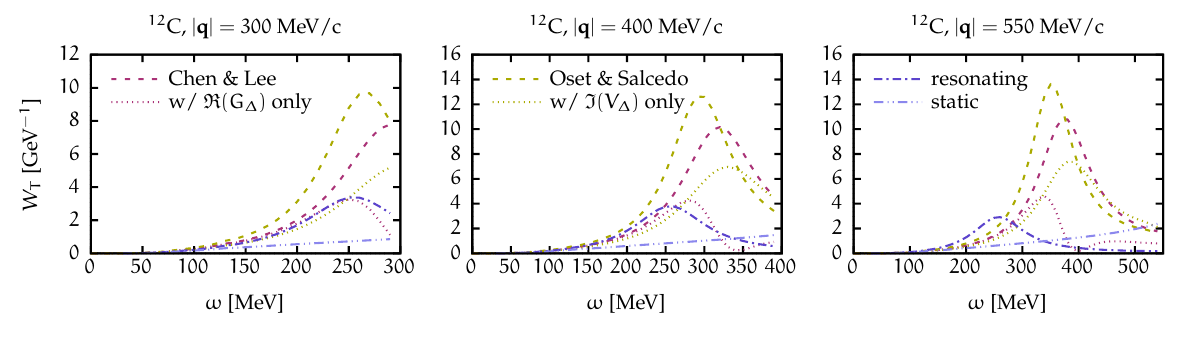}
  \caption{Energy dependence of 2p2h hadronic $^{12}\mathrm{C}(e,e')$ responses induced by $\Delta$-mediated meson exchange currents for fixed three-momentum transfer. The purple dashed and dotted lines present the calculation using the $\Delta$-resonance in-medium properties of Ref.~\cite{Chen:1988xq}, with the latter providing results with the resonance propagator truncated to its real part only. The olive dashed and dotted lines present the calculation using the $\Delta$-resonance in-medium properties of Ref.~\cite{Oset:1987re}, with the latter providing results with only the imaginary part of the resonance self-energy. The iris dash-dot and lavender dash-dot-dot show the results of applying the resonating approximation of Ref.~\cite{VanderSluys:1995rp} and the static approximation of Ref.~\cite{Ryckebusch:1993tf}, respectively.}
  \label{fig:incl_q_2p2h_del}
\end{figure*}

Meson-exchange currents provide visible contributions to inclusive scattering, leading to both 1p1h and 2p2h final states. In Fig.~\ref{fig:incl_q_corr2_mecs}, we present the absolute and fractional corrections to the 1p1h transverse hadronic response. The top row shows the absolute correction, defined as the difference between the calculation including two-body currents and the one obtained within the impulse approximation, $\Delta W_{T} := W_{T}^{\text{IA+MEC}}-W_{T}^{\text{IA}}$, while the bottom row displays the corresponding fractional contribution, $\Delta W_{T} / W_{T}^{\text{IA+MEC}}$. The results are shown for the kinematical regimes introduced in Fig.~\ref{fig:resp_pwdw}. We observe that the seagull and pion-in-flight terms contribute with opposite signs, resulting in a small overall correction when combined. Conversely, $\Delta$-currents provide a significant enhancement of the interaction strength and dictate the overall positive impact of the MECs on the 1p1h strength. Relative to the total response, the fractional contribution of all MEC terms grows as one approaches the dip region and decreases with increasing $Q^2$.

It is important to emphasize that within the 1p1h sector, the contribution of the pure two-body term $|\hat{J}^{[2]}|^2$ is close to negligible. Consequently, the observed corrections attributed to meson-exchange currents are almost entirely driven by the interference term $2\Re(\hat{J}^{[1]\dagger}\hat{J}^{[2]})$. This explains why the sign of MEC components is so critical: while the 2p2h cross section---dominated by the $|\hat{J}^{[2]}|^2$ term---is insensitive to relative phases between the different terms in the current operator, the 1p1h response is defined by the constructive or destructive nature of this interference. We find that the interference contributions from our seagull and pion-in-flight terms have the opposite sign compared to those reported in Refs.~\cite{Franco-Munoz:2022jcl,TaniaPriv}. Furthermore, both our model and Ref.~\cite{Franco-Munoz:2022jcl} exhibit an opposite sign in the $\Delta$-currents interference compared to the results of Ref.~\cite{Amaro:1994fx} and the subsequent variations of that framework. While the origin of this sign flip relative to various models in the literature is attributed to a different choice of the arbitrary phase, the former sign differences remain a subject of investigation, as they persist even when consistent constants and current operators are employed. Such discrepancies may point to subtle sensitivities in the interference between the impulse approximation and the two-body currents, or to the different treatment of the nuclear wave functions between models.

In Figs.~\ref{fig:incl_q_2p2h_mecs} and~\ref{fig:incl_q_2p2h_del}, we present the 2p2h transverse hadronic responses. Similarly to the 1p1h sector, the destructive interference between the non-resonant MECs significantly reduces their impact on the predicted responses. At higher energy transfers $\omega$, we observe a pronounced $\Delta$-resonance peak, with the seagull and pion-in-flight contributions being comparable in magnitude for $\omega \lesssim 200 \ \mathrm{MeV}$.

\begin{figure*}[ht!]
  \centering
  \includegraphics[width=1.0\textwidth]{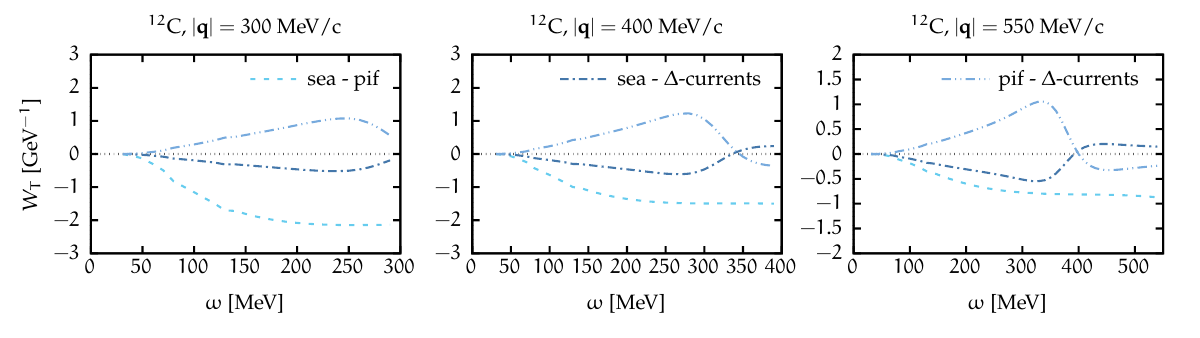}
  \caption{Energy dependence of the interference between pairs of meson-exchange currents types in 2p2h hadronic $^{12}\mathrm{C}(e,e')$ responses for fixed three-momentum. The cyan dashed, blue dash-dot-dot, and light blue dashed lines present the seagull and pion-in-flight, seagull and $\Delta$-mediated, and pion-in-flight and $\Delta$-mediated MEC pairs, respectively.}
  \label{fig:incl_q_2p2h_mecs_int}
\end{figure*}

As extensively discussed in the literature, the predicted strength of the $\Delta$-currents is sensitive to the choices for the $\Delta$ propagators. We observe that the static approximation yields little strength at the presented kinematics, though it is known to diverge unphysically at higher energy transfers. The resonating approach, while successful in stabilizing this large-$\omega$ behavior, fails to accurately predict the position of the $\Delta$-resonance peak. Ultimately, our results suggest that utilizing the full propagator prescription is essential to reproduce the correct magnitude and shape of the 2p2h response. We find that the ratio between the full calculation and static approximation is consistent with the results of Refs.~\cite{Dekker:1994yc} and~\cite{DePace:2003spn}. However, we note that the latter finds that the static model begins to diverge at lower values of $\omega$ than observed here. Our results are consistent with those of J.E.~Amaro \textit{et al.}~\cite{Amaro:1994fx}.

As presented in Fig.~\ref{fig:incl_q_2p2h_del}, our results exhibit a high degree of sensitivity to the choice of the in-medium potential $V_\Delta$. When the real part of the self-energy is omitted, the $\Delta$-resonance peak shifts toward excessively higher energy transfers. Introducing the complete density-dependent potential of E.~Oset and L.~Salcedo~\cite{Oset:1987re} shifts the peak back toward lower energies. However, this is accompanied by an increased interaction strength and a sharper resonance peak. Our default choice, the constant potential of C.~Chen and T.~Lee~\cite{Chen:1988xq}, provides a middle ground between these prescriptions, yielding a balanced peak position and width.

Furthermore, we investigated the implications of the $\Re(G_\Delta)$-only prescription discussed in Sec.~\ref{subsec:mec}. Omitting the imaginary part of the propagator reduces the predicted strength at the peak by a factor of about 2.5 and produces an unphysical energy dependence of $W_T$, with the peak dropping sharply to zero at resonance, as anticipated above (see also Ref.~\cite{Niewczas:2023phd}).

To complete our analysis of the meson-exchange currents, we examine the interference patterns between the individual current components in the 2p2h response, as presented in Fig.~\ref{fig:incl_q_2p2h_mecs_int}. We find that the seagull-$\Delta$ and pion-in-flight-$\Delta$ interferences exhibit a striking ``mirror'' symmetry: the former provides a mostly negative contribution while the latter is mostly positive, with both displaying nearly identical shapes. This suggests that the $\Delta$ resonance couples to these gauge-linked components in a fundamentally similar manner, differing only by a relative sign. In contrast, the sea-pif interference is found to be significantly stronger and explicitly negative across the entire kinematics range. This dominant destructive interference between the non-resonant terms suppresses their combined impact, further reinforcing the $\Delta$-isobar current as the primary driver of the 2p2h strength.

\subsection{Short-range correlations}
\label{subsec:src}

\begin{figure}[ht!]
  \centering
  \includegraphics[width=0.48\textwidth]{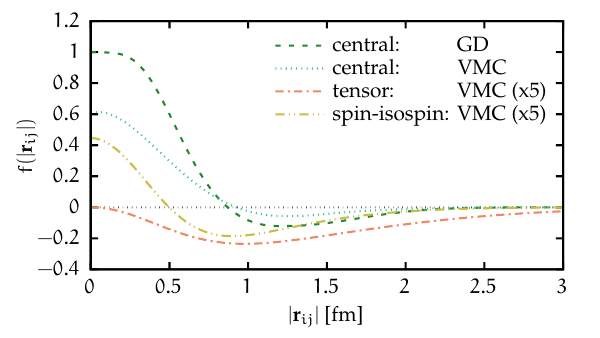}
  \caption{Correlation functions used in the presented calculations. The green dashed line is the central correlation function $f_c(|\mathbf{r}_{ij}|)$ calculation by C.~Gearhart \textit{et al.}~\cite{Gearhart:1994phd}. The mint dashed, orange dash-dot, and yellow dash-dot-dot lines provide the central $f_c(|\mathbf{r}_{ij}|)$, spin-isospin $f_{\sigma\tau}(|\mathbf{r}_{ij}|)$, and tensor $f_{t\tau}(|\mathbf{r}_{ij}|)$ functions of Ref.~\cite{Pieper:1992gr}, respectively.}
  \label{fig:correlations}
\end{figure}

In independent-particle models, such as the one described in Subsection~\ref{subsec:impulse_approximation}, the nuclear many-body state is represented as a Slater determinant $| \, \Phi \, \rangle$. Our HF is thus a model for nuclear wave functions at low resolution scales. To restore the high-resolution SRC physics integrated out of this picture~\cite{Tropiano:2021qgf}, we employ a many-body correlation operator $\hat{\mathcal{G}}$ acting on the independent-particle state~\cite{Ryckebusch:2019oya}:
\begin{equation}
\label{eq:src_wavef}
    | \, \Psi \, \rangle = \frac{1}{\sqrt{\mathcal{N}}} \hat{\mathcal{G}} | \, \Phi \, \rangle ,
\end{equation}
where $\mathcal{N} = \langle \, \Phi \, | \hat{\mathcal{G}}^\dagger\hat{\mathcal{G}} | \, \Phi \, \rangle$ is the normalization constant. We approximate the correlation operator using a cluster expansion, truncated at the two-body level:
\begin{equation}
\label{eq:correlation_operator}
    \hat{\mathcal{G}} \simeq \hat{\mathcal{S}} \left( \prod_{i<j=1}^A \left[ 1 + \hat{l}_{ij} \right] \right) ,
\end{equation}
where $\hat{\mathcal{S}}$ is the symmetrization operator. The two-body correlation operator $\hat{l}_{ij}$ is decomposed as
\begin{equation}
\begin{split}
\label{eq:src_operator}
    \hat{l}_{ij} = & -f_c(|\mathbf{r}_{ij}|) \\
    & + f_{t\tau} (|\mathbf{r}_{ij}|)\hat{S}_{ij}(\bm{\tau}_i\cdot\bm{\tau}_j) \\
    & + f_{\sigma\tau}(|\mathbf{r}_{ij}|) (\bm{\sigma}_i\cdot\bm{\sigma}_j)(\bm{\tau}_i\cdot\bm{\tau}_j) ,
\end{split}
\end{equation}
with the tensor operator $\hat{S}_{ij}$ defined in Eq.~\eqref{eq:tensor_operator}.

The spatial dependence of the correlation operator is governed by the universal ($A$-independent) functions $f_{x\in \{c,t\tau,\sigma\tau\}}(|\mathbf{r}_{ij}|)$, which are typically extracted from many-body computations. In this work, we utilize the central correlation function obtained by C.~Gearhart and W.~Dickhoff (GD)~\cite{Gearhart:1994phd} alongside the central, tensor, and spin-isospin functions from the Variational Monte Carlo (VMC) calculations of S.C.~Pieper \textit{et al.}~\cite{Pieper:1992gr}. The GD function was derived using Green's function methods for nuclear matter with the Reid potential~\cite{Reid:1968sq}, while the VMC functions were obtained for finite nuclei using the Argonne $v_{14}$ potential~\cite{Wiringa:1984tg} and the Urbana VII three-nucleon interaction~\cite{Schiavilla:1985gb}.

As shown in Fig.~\ref{fig:correlations}, the central correlation functions exhibit a ``hard core'' at short separation ($|\mathbf{r}_{ij}| < 1 \ \mathrm{fm}$), reflecting the strong repulsion of the $NN$ potential. In contrast, the tensor and spin-isospin correlations, while smaller in magnitude at the origin, extend over larger internucleon distances. Comparisons with exclusive $(e,e'pp)$ data, which are highly sensitive to short-range dynamics, show a preference for the intermediate-strength GD function over the relatively ``soft'' VMC central parametrization~\cite{Blomqvist:1998gq}. However, because the VMC framework provides a robust description of the long-range tensor and spin-isospin structures, unless otherwise specified, we adopt a hybrid choice combining the GD central function with the VMC tensor and spin-isospin functions.

\begin{figure}[t]
    \centering 
    \subfloat[]{\includegraphics[width=0.19\textwidth]{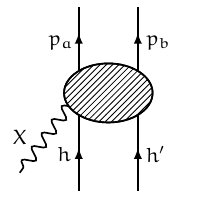}}
    \subfloat[]{\includegraphics[width=0.19\textwidth]{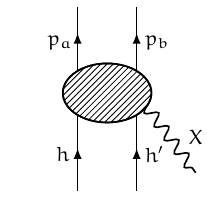}}
    \caption{Graphical representation of the diagrams involved in the two-nucleon knock-out processes induced with the short-range correlations dynamics.} 
    \label{fig:diagrams_2p2h_src} 
\end{figure}

This framework, also dubbed the low-order correlation operator approximation (LCA), has been applied across mass range $4 \leq A \leq 208$ and proved able to reproduce the aggregated effect of SRC contributions to $A(e,e')$ responses~\cite{Ryckebusch:2019oya}. Additionally, LCA, which retains terms linear and quadratic in the correlation functions, provides nuclear momentum distributions for light nuclei that are in very good agreement to those obtained from ab-initio Quantum Monte Carlo by the Argonne group using the $v_{18}$ nucleon-nucleon interaction~\cite{Wiringa:2013ala,Lonardoni:2017egu}.

\subsubsection{Effective nuclear current}

Short-range correlations can be incorporated into numerical calculations by introducing effective nuclear current operators. A transition matrix element between correlated ground and final states is expressed as
\begin{widetext}
\begin{equation}
    \langle \, \Psi_B \, | \, \hat{J}_{\mu}(\mathbf{q}) \, | \, \Psi_A \, \rangle
    = \frac{1}{\sqrt{\mathcal{N}_A\mathcal{N}_B}} \langle \, \Phi_B \, | \, \hat{\mathcal{G}}^\dagger \hat{J}_{\mu}(\mathbf{q}) \hat{\mathcal{G}} \, | \, \Phi_A \, \rangle
    = \frac{1}{\sqrt{\mathcal{N}_A\mathcal{N}_B}} \langle \, \Phi_B \, | \, \hat{J}_{\mu}^\mathrm{eff}(\mathbf{q}) \, | \, \Phi_A \, \rangle .
\end{equation}
\end{widetext}
with the effective nuclear current defined by the cluster expansion~\cite{Ryckebusch:1997gn}:
\begin{equation}
\label{eq:src:current_eff}
    \hat{J}_{\mu}^\mathrm{eff}(\mathbf{q}) =
    \prod_{i<j=1}^A \left[ 1 + \hat{l}_{ij} \right]^\dagger
    \hat{J}_{\mu}
    \prod_{k<l=1}^A \left[ 1 + \hat{l}_{kl} \right] .
\end{equation}

\begin{figure*}[ht!]
  \centering
  \includegraphics[width=1.0\textwidth]{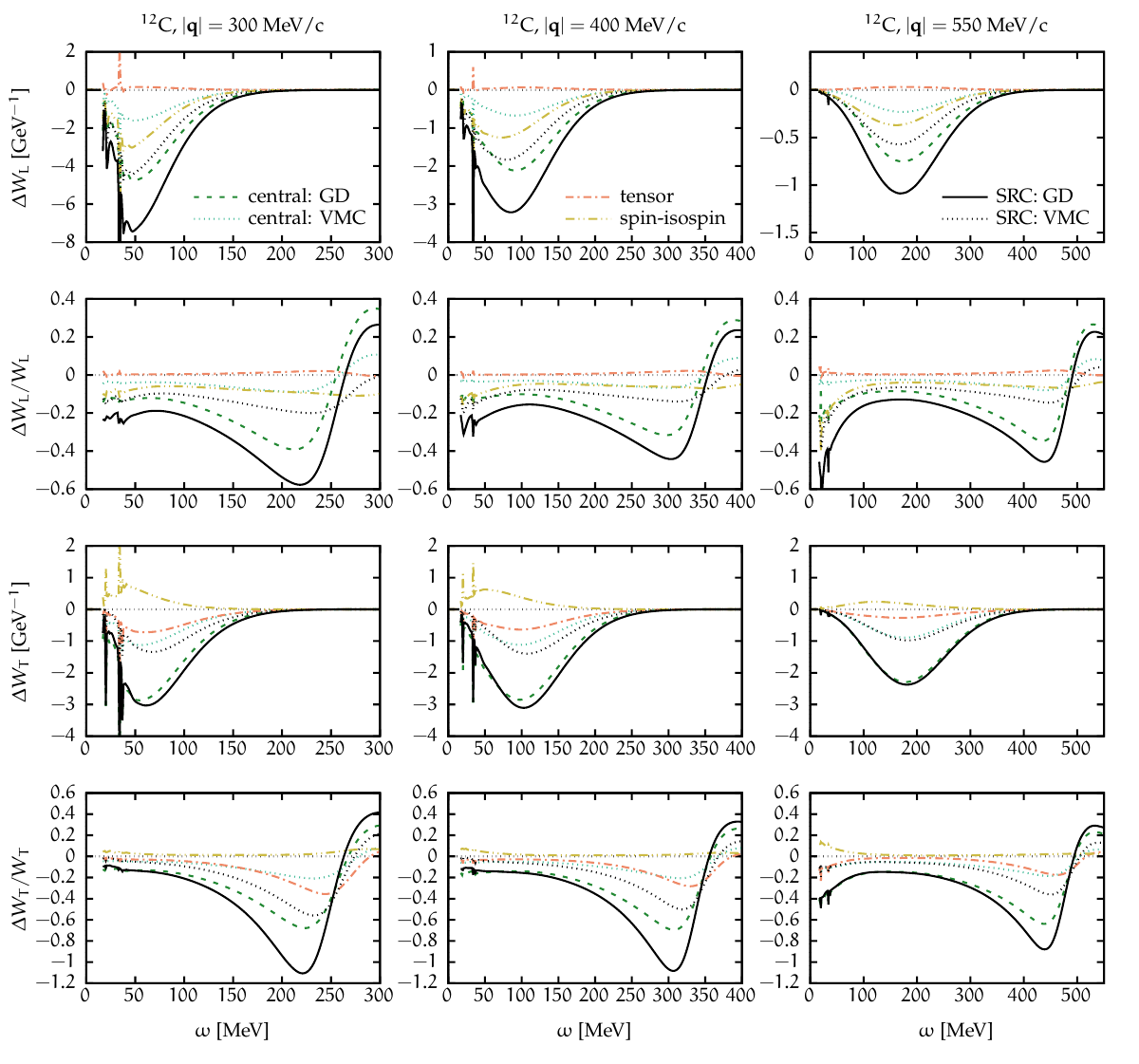}
  \caption{Energy dependence of the absolute and fractional SRC corrections to the 1p1h hadronic $^{12}\mathrm{C}(e,e')$ responses for fixed three-momentum transfer. The green dashed line presents the effects of central correlations, while the solid black line---the total short-range correlations contribution, both using the central correlation GD function of Ref.~\cite{Gearhart:1994phd}. The mint dashed, orange dash-dot, yellow dash-dot-dot, and black dotted lines provide the central, spin-isospin, tensor, and total correlation effects using the VMC functions of Ref.~\cite{Pieper:1992gr}, respectively.}
  \label{fig:incl_q_corr_srcs}
\end{figure*}

\begin{figure*}[t]
  \centering
  \includegraphics[width=1.0\textwidth]{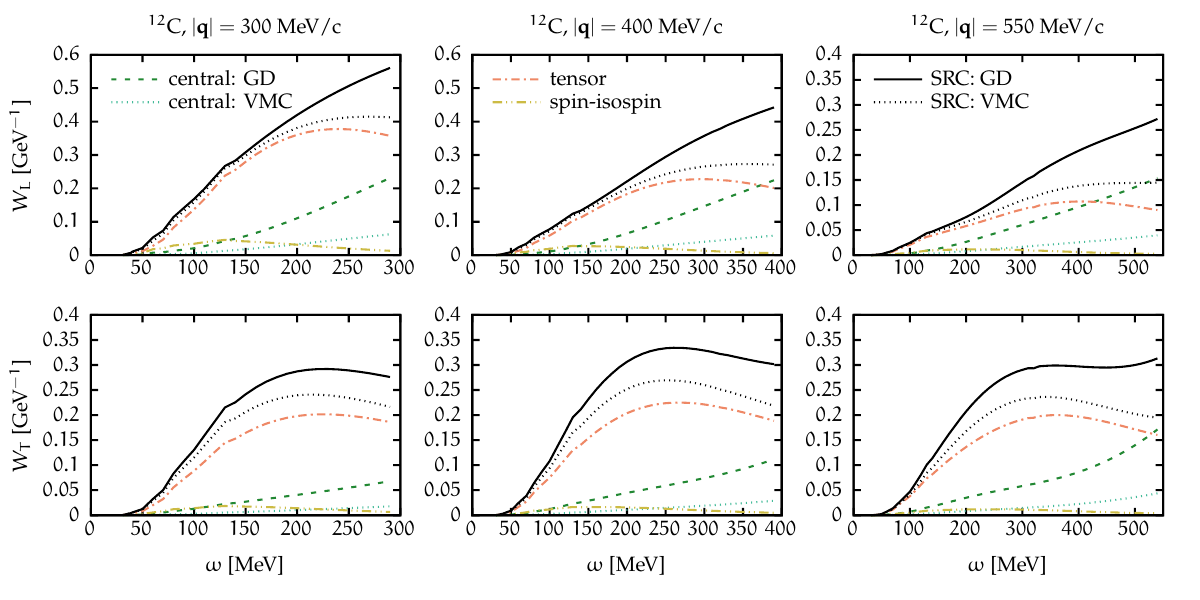}
  \caption{Energy dependence of 2p2h hadronic $^{12}\mathrm{C}(e,e')$ responses induced by short-range correlations for fixed three-momentum transfer. Line conventions as in Fig.~\ref{fig:incl_q_corr_srcs}.}
  \label{fig:incl_q_2p2h_srcs}
\end{figure*}

Because SRCs require nucleons to be spatially close, and the probability of three or more nucleons simultaneously occupying such a close configuration is small, two-body correlations constitute the dominant SRC-induced dynamics. Two approximations follow from this. First, we use a cluster expansion of Eq.~\eqref{eq:src:current_eff} and retain terms linear in the correlation operator $\hat{l}_{ij}$. Second, since we perform calculations leading to one or two final state nucleons in the continuum (Eqs.~\eqref{eq:1p1h_state} and~\eqref{eq:2p2h_state}), we refrain from including terms expressing correlations between occupied and asymptotically free single-particle states~\cite{Ryckebusch:1995usx}. Incorporating the interaction dynamics containing one-body and MEC components, $\hat{J}_{\mu} \simeq \sum_{i=1}^{A} \hat{J}_{\mu,i}^{[1]}+\sum_{i<j}^{A} \hat{J}_{\mu,ij}^{[2],\text{MEC}}$, we decompose the effective nuclear current into:
\begin{equation}
\begin{split}
    \hat{J}_{\mu}^{\textrm{eff}}(\mathbf{q}) \simeq
    \sum_{i=1}^{A} \hat{J}_{\mu,i}^{[1]}(\mathbf{q})
    & + \sum_{i<j}^{A} \hat{J}_{\mu,ij}^{[2],\textrm{SRC}}(\mathbf{q}) \\
    & + \sum_{i<j}^{A} \hat{J}_{\mu,ij}^{[2],\textrm{MEC}}(\mathbf{q}) ,
\label{eq:mec-src}
\end{split}
\end{equation}
where the SRC-induced current is defined as:
\begin{equation}
    \hat{J}_{\mu,ij}^{[2],\mathrm{SRC}}(\mathbf{q}) = \left[ \hat{J}_{\mu,i}^{[1]}(\mathbf{q}) + \hat{J}_{\mu,j}^{[1]}(\mathbf{q}) \right] \hat{l}_{ij} .
\end{equation}
We identify two distinct two-body operators: $\hat{J}_{\mu,ij}^{[2],\textrm{SRC}}$, which modifies the $\gamma^\ast A$ interaction due to dynamically generated nucleon-nucleon correlations beyond the IPM, and $\hat{J}_{\mu,ij}^{[2],\textrm{MEC}}$, which corrects the $\gamma^\ast A$ coupling via $\pi$ and $\Delta$ degrees of freedom. As the short distance (high-momentum) behavior of the meson-exchange currents is regulated by the form factors of Eq.~\eqref{eq:monoform}, we refrain from introducing higher-order corrections that stem from combining MEC with the SRC correlation operator of Eq.~\eqref{eq:correlation_operator}.

Physically, the expansion in Eq.~\eqref{eq:mec-src} captures two distinct phenomena. First, it accounts for the redistribution of single-particle strength caused by high-momentum components in the ground state---an effect typically modeled by introducing spectroscopic factors. Here, this microscopic effect is generated dynamically, describing the partial occupancy of shell-model states and the characteristic high-momentum tail of nuclear momentum distributions that cannot be accommodated within the IPM. Second, the direct coupling of the electromagnetic probe to a nucleon pair via $\hat{J}_{\mu,ij}^{[2],\mathrm{SRC}}$ induces explicit 2p2h final states. As illustrated schematically in Fig.~\ref{fig:diagrams_2p2h_src}, this framework incorporates the effects of short-range correlations on an equal footing with other mechanisms while utilizing well-established computational techniques to deal with the multidimensional integrals. By representing the short-range correlation as an effective vertex---distinct from the explicit pion exchange in the MECs---one can systematically evaluate the sensitivity of the interaction strength to SRCs relative to other modeled dynamics.

\subsubsection{Hadronic responses}

\begin{figure*}[ht!]
  \centering
  \includegraphics[width=1.0\textwidth]{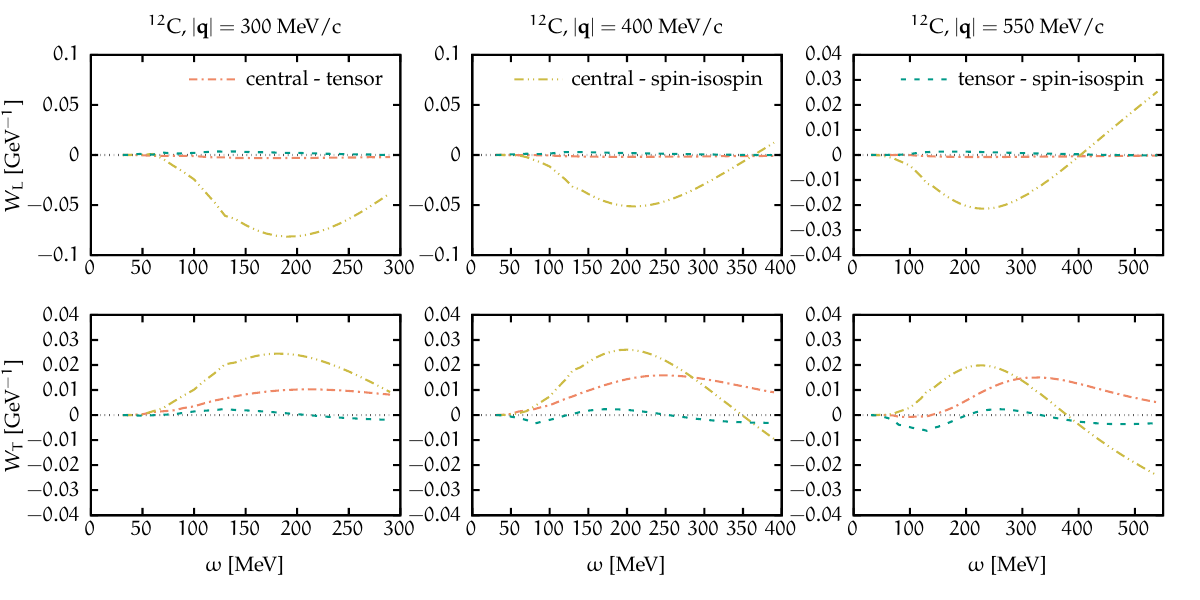}
  \caption{Energy dependence of the interference between pairs of short-range correlation types in 2p2h hadronic $^{12}\mathrm{C}(e,e')$ responses for fixed three-momentum. The orange dash-dot, yellow dash-dot-dot, and teal dashed lines present the central and tensor, central and spin-isospin, and tensor and spin-isospin pairs, respectively.}
  \label{fig:incl_q_2p2h_srcs_int}
\end{figure*}

\begin{figure*}[ht!]
  \centering
  \includegraphics[width=1.0\textwidth]{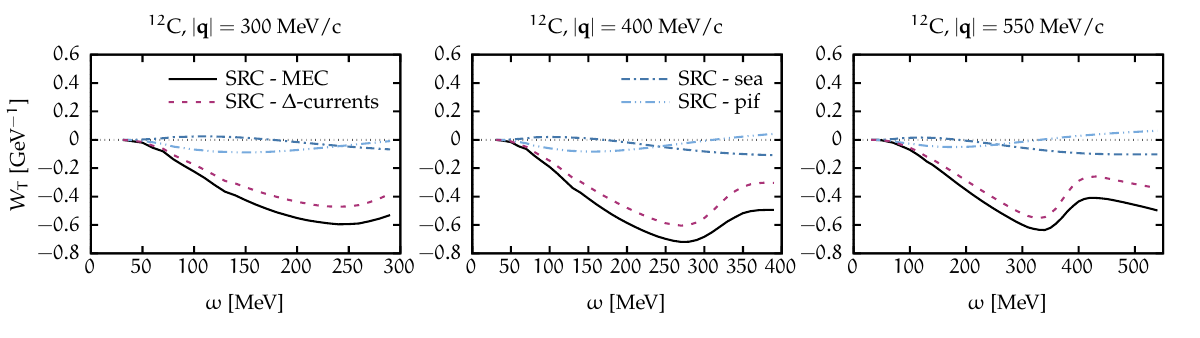}
  \caption{Energy dependence of the interference between the particular meson-exchange currents and total short-range correlations in 2p2h hadronic $^{12}\mathrm{C}(e,e')$ responses for fixed three-momentum values. The black solid, purple dashed, blue dash-dot, and light blue dash-dot-dot present the MEC and SRC, $\Delta$-currents and SRC, seagull currents and SRC, and pion-in-flight currents and SRC interferences, respectively.}
  \label{fig:incl_q_2p2h_all_int}
\end{figure*}

Hadronic responses provide insight into the impact of effective short-range correlation currents on the strength of the electromagnetic interaction. In contrast to MECs, SRCs contribute to both longitudinal and transverse responses. Fig.~\ref{fig:incl_q_corr_srcs} illustrates the absolute corrections to the 1p1h IA responses due to SRCs, defined as $\Delta W_{L,T} := W_{L,T}^{\text{IA+SRC}}-W_{L,T}^{\text{IA}}$, as well as the fractional contributions, $\Delta W_{L,T} / W_{L,T}^{\text{IA+SRC}}$. Consistent with the expectation that correlations deplete single-particle strength by shifting it to higher momenta and energies, the overall impact of short-range correlations is a reduction of both the longitudinal and transverse responses. Our current results align with the physical requirement that SRCs reduce the occupancy of mean-field states.

The magnitude of this suppression is highly sensitive to the choice of central correlation function; the ``hard core'' GD parametrization yields a more pronounced reduction compared to the ``softer'' VMC version, bringing the longitudinal results into closer agreement with data. Interestingly, while most components contribute to this reduction, the tensor correlation in the longitudinal response and the spin-isospin correlation in the transverse response provide small positive contributions. Overall, the reduction of $W_L$ and $W_T$ due to SRCs is significant, particularly at lower energy transfers where it provides a microscopic foundation for the quenching effect traditionally associated with empirical spectroscopic factors.

In Fig.~\ref{fig:incl_q_2p2h_srcs}, we present the impact of short-range correlations on the 2p2h hadronic responses. We observe that the tensor component is the dominant contributor to the SRC strength, particularly at lower values of $\omega$, which is consistent with the predominant role of the tensor force in generating high-momentum pairs. However, as the energy transfer $\omega$ increases, the central correlation contribution becomes the primary driver of the response, while the spin-isospin contribution remains negligible. This indicates that while SRCs provide most of the two-nucleon knock-out strength at high $Q^2$, their contribution to the $^{12}\mathrm{C}(e,e')$ cross sections at low $Q^2$ is relatively modest.

Interferences among two-body current dynamics provide valuable insight into the underlying structure of our model. In Fig.~\ref{fig:incl_q_2p2h_srcs_int}, we display the interference contributions between individual SRC correlation types. In the longitudinal channel, the only non-negligible interference occurs between the central and spin-isospin terms. In the transverse channel, we observe additional interferences of the central-(spin-isospin) and central-tensor contributions. These effects are largely cosmetic, as their magnitudes are significantly smaller than the interferences found between MEC components, suggesting that the individual correlation channels act largely independently within the 2p2h sector.

\begin{figure*}[ht!]
  \centering
  \includegraphics[width=1.0\textwidth]{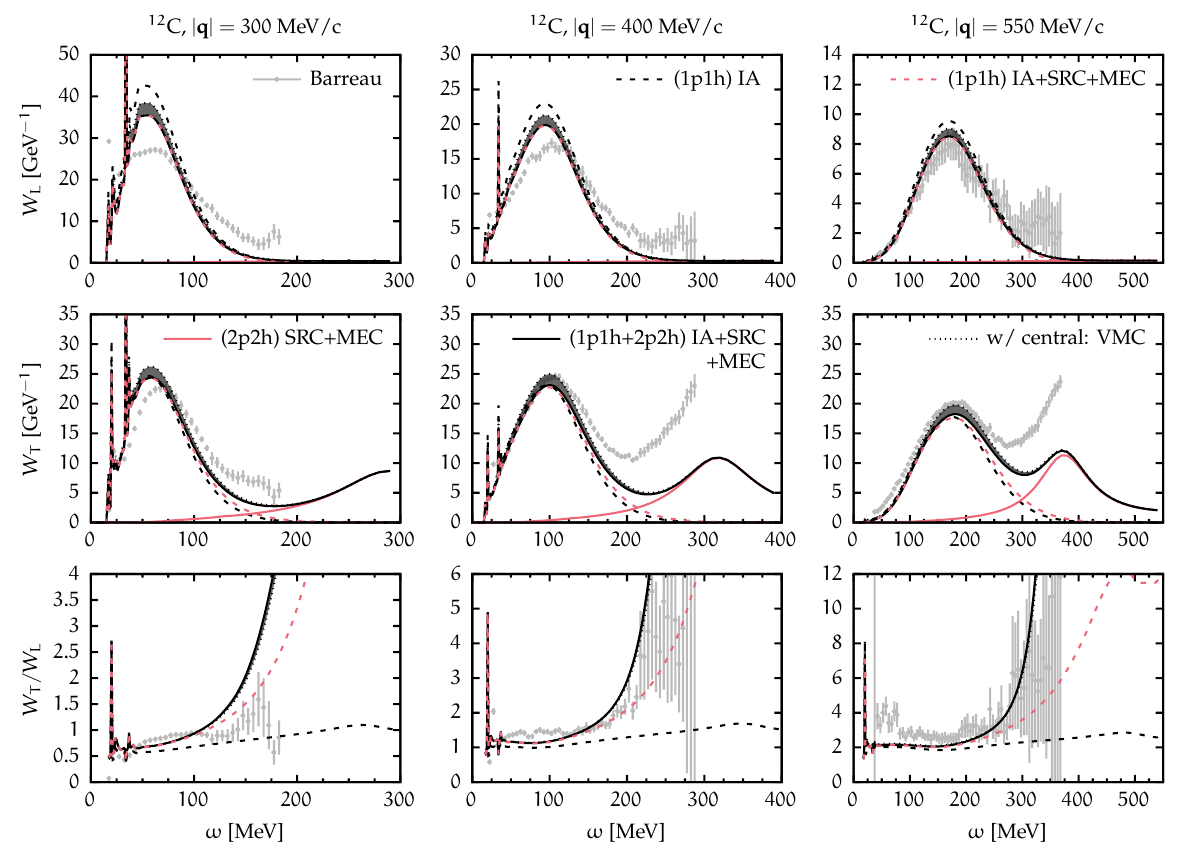}
  \caption{Energy dependence of the inclusive hadronic responses and their ratio for fixed three-momentum transfer values in $^{12}\mathrm{C}(e,e')$ compared with the experimental results of Ref.~\cite{Barreau:1983ht}. The black dashed lines present the impulse approximation calculation, while the red dashed lines include the effects of two-body currents. The red solid lines provide the 2p2h responses using the $\Delta$-resonance in-medium properties of Ref.~\cite{Chen:1988xq}. The black solid lines are the incoherent sums of one- and two-nucleon knock-out responses. The shaded area, bounded by the black solid and dotted lines indicates the sensitivity of the total response to the choice of the central correlation function, between GD and VMC parametrizations. For clarity, the ratio of isolated 2p2h responses is omitted from the bottom row.}
  \label{fig:incl_q_resp_all:barreau}
\end{figure*}

\begin{figure*}[ht!]
  \centering
  \includegraphics[width=1.0\textwidth]{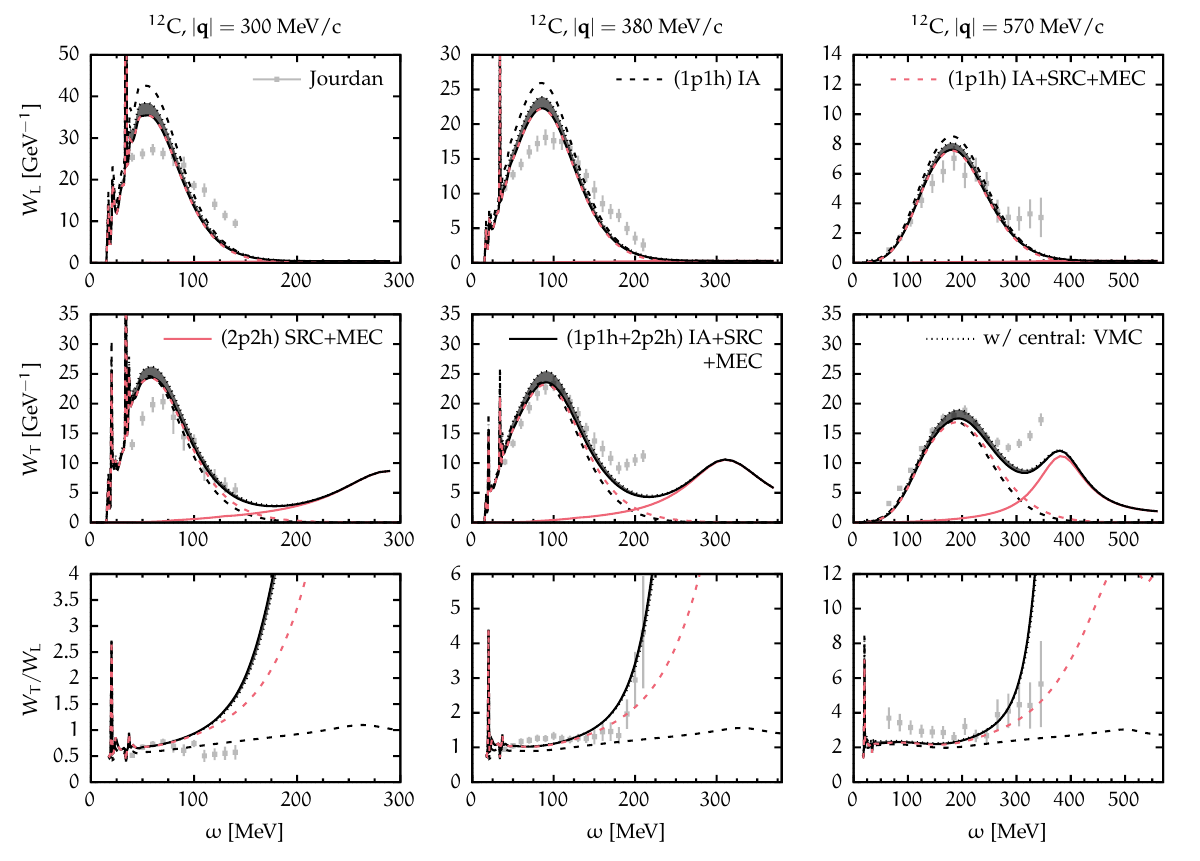}
  \caption{Energy dependence of the inclusive hadronic responses and their ratio for fixed three-momentum transfer values in $^{12}\mathrm{C}(e,e')$ compared with the experimental results of Ref.~\cite{Jourdan:1996np}. Line conventions as in Fig.~\ref{fig:incl_q_resp_all:barreau}.}
  \label{fig:incl_q_resp_all:jourdan}
\end{figure*}

Finally, Fig.~\ref{fig:incl_q_2p2h_all_int} examines the interference between the total SRC current and individual MEC components. While the interferences with the seagull and pion-in-flight currents are small and exhibit sign changes that complicate physical interpretation, the interference between the SRC and $\Delta$-currents is quite substantial and destructive. This effect essentially ``cannibalizes'' the strength provided by the 2p2h SRCs, leading to a 2p2h response with the transverse channel dominated by $\Delta$-currents.

It is instructive to compare these findings with early studies on two-body current interferences, notably those by W.~Alberico \textit{et al.}~\cite{Alberico:1983zg}, and K.~Shimizu and A.~Faessler~\cite{Shimizu:1980kb}. The former reported negative and positive interferences between the correlation currents and the seagull and pion-in-flight currents, respectively. Additionally, both references reported positive interference between the correlation currents and $\Delta$-current contributions. These interference signs stand in direct opposition to the findings of the present work. Such discrepancies likely arise from the differing implementations of correlations; while earlier works often treated correlations via specific loop corrections to the current, our SRC framework is implemented as an effective operator acting on a mean-field ground state. As the earlier literature uses wave functions different from ours, a direct one-to-one comparison of the model results is not possible. Nevertheless, the dominance of the $\Delta$-current in the transverse response remains a robust and universal feature in these different modeling approaches.
\section{Results and discussion}
\label{sec:results}

The model presented in this work provides results for one- and two-body nuclear current contributions to both 1p1h and 2p2h final states in $^{12}\mathrm{C}(e,e')$ scattering at intermediate energies. Having analyzed all components of the model, we now combine the one-body impulse approximation with the effective SRC currents and pionic MECs to evaluate the total inclusive response of $^{12}\mathrm{C}$. Figs.~\ref{fig:incl_q_resp_all:barreau} and~\ref{fig:incl_q_resp_all:jourdan} confront our full calculations with data.

The net effect of our SRC+MEC implementation on the 1p1h sector is a $|\mathbf{q}|$-dependent reduction in the longitudinal response and a small $|\mathbf{q}|$-independent enhancement of the transverse strength. As discussed in Section~\ref{subsec:src}, the suppression of the longitudinal strength is primarily governed by the central correlation function. The use of the GD function---successfully applied previously to exclusive and semi-exclusive processes~\cite{Janssen:1999xy,Blomqvist:1998gq}---is essential here. Compared to our previous studies, we now reach a higher degree of numerical stability. Despite these improvements, our results overestimate the longitudinal strength at low $|\mathbf{q}|$ and $\omega$. This persistent discrepancy may be attributed to the fact that our independent-particle basis lacks long-range correlations to describe the collective behavior at low $Q^2$. Incorporating a formalism such as CRPA~\cite{Jachowicz:2021ieb} appears necessary to complete the description of the collective nuclear response at low energy transfers. However, combining the two frameworks is conceptually not trivial and technically challenging.

The sensitivity of our results to the choice of the central correlation function---represented by the shaded areas in Figs.~\ref{fig:incl_q_resp_all:barreau} and~\ref{fig:incl_q_resp_all:jourdan}---provides further insight into the interplay between different correlation scales. In principle, a ``harder'' central correlation function, such as the GD parametrization, removes more 1p1h strength and provides more 2p2h strength. However, within our framework, the 2p2h sector is dominated by $\Delta$-currents and tensor SRCs, so the primary impact of the central correlation choice is felt in the quenching of the 1p1h channel.

In the 2p2h sector, the response exhibits a clear evolution in dynamics. In line with the conclusions drawn in other studies, the longitudinal response is largely unaffected by the dynamics of the two-body currents. The transverse response at lower energy transfers originates from SRCs and non-resonant MECs, becoming quickly dominated by the $\Delta$-currents as $\omega$ approaches the $\Delta$ peak. Nevertheless, we slightly underestimate the strength in the ``dip'' region between the quasielastic and $\Delta$ peaks. While adding 2p2h strength brings the predictions closer to the data, the results remain highly sensitive to the in-medium properties of the $\Delta$-isobar. Specifically, the choices made in the $\Delta$-propagator and its self-energy (Fig.~\ref{fig:delta_inmedium}) represent a major source of uncertainty.

Some of the dip-region strength would in principle also be accounted for by an explicit treatment of single-pion production. As noted in Ref.~\cite{Niewczas:2023phd}, clearly distinguishing between virtual pion exchange (MEC) and real pion production (SPP) is non-trivial from a conceptual point of view. While some modeling approaches attempt to avoid double-counting by isolating specific channels, we argue that this strength is shared between 2p2h and SPP. Disentangling their relative contributions is an open question, as, to our knowledge, no model-independent prescription to separate the FSI-corrected SPP and MEC processes exists.

The dominance of the 2p2h strength by the $\Delta$-current, with a subsequent influence of the interference between one- and two-body currents in the 1p1h channel, provides arguments for modeling the inclusive 1p1h sector with effective spectroscopic factors and averaged treatment of nucleon-nucleon correlations~\cite{Benhar:1994hw,Gonzalez-Jimenez:2019ejf}. However, as presented in Fig.~\ref{fig:resp_pwdw}, such approximations cannot succeed without the proper treatment of nucleon distortion and implicit Pauli blocking, especially when dealing with more exclusive final states.

Finally, we address the ratio of transverse-to-longitudinal ($W_T/W_L$) responses. While our model qualitatively captures the trend of the data, the quantitative differences in $W_T / W_L$ point to a remaining imbalance in the simultaneous description of $W_T$ and $W_L$. Throughout the QE peak, our calculation tends to underestimate the ratio. At lower $|\mathbf{q}|$, this follows directly from the longitudinal overestimation discussed above, which is not compensated by a corresponding excess in $W_T$. At higher $|\mathbf{q}|$, however, $W_T$ falls below the data. In our calculation, the SRC-induced quenching acts equally on both responses and diminishes with $|\mathbf{q}|$, pointing to some deficiency in the modeling of the transverse strength.

The ratio tends to be overestimated in the dip region. The two-body corrections to the 1p1h channel already bring it close to the data, but the inclusion of 2p2h final states pushes it above the experimental values. Part of this excess likely reflects the non-trivial overlap between virtual pion exchange and single-pion production discussed above, which makes a clean identification of the underlying reaction mechanisms difficult in this regime. More broadly, achieving a simultaneous description of both responses remains a delicate balancing act among short-range dynamics, pionic degrees of freedom, and collective nuclear effects.

\section{Conclusions}
\label{sec:conclusions}

The era of precision neutrino physics demands lower systematic uncertainties in predictions for neutrino-nucleus cross sections. This motivates a shift from legacy modeling techniques to microscopic models that can be rigorously validated against electron-scattering data. In this work, we presented a comprehensive calculation of the inclusive $^{12}\mathrm{C}(e,e')$ responses within a framework that treats one- and two-body currents alongside short-range correlations and meson-exchange currents. By utilizing a non-relativistic nuclear mean-field potential for both initial and final hadronic states, we properly treat nucleon distortion and Pauli blocking, and thus avoid spurious non-orthogonality contributions to the matrix elements. These effects are vital to a realistic description of the nuclear response.

Our analysis demonstrates that the observed features of the nuclear response arise from a delicate interplay between different microscopic mechanisms. Specifically, the inclusion of SRCs provides a quenching of the one-body strength, whereas MECs supply the localized transverse enhancement required to describe the dip region and the $\Delta$-peak. The high degree of consistency achieved by combining SRCs and MECs allows for a more natural resolution of the transverse-to-longitudinal response ratio than is possible in Fermi gas or simple independent-particle models.

However, the persistent differences observed at low energy transfers highlight that short-range dynamics are only one piece of the puzzle. The overestimation of the longitudinal peak suggests that long-range correlations should be incorporated into this framework, a task that remains far from trivial. Furthermore, the sensitivity of the 2p2h sector to in-medium $\Delta$ properties, and the potential overlap with single-pion production channels remain the primary frontiers for future refinements.

Ultimately, there is a fundamental difference between a model that merely reproduces experimental data through parameter tuning and a microscopic description that offers genuine physical insight. By constraining the vector sector through the lens of electron scattering, this work provides a grounded foundation for the axial-current extensions required by upcoming experiments like DUNE~\cite{DUNE:2020lwj} and Hyper-Kamiokande~\cite{Hyper-Kamiokande:2018ofw}. As the field moves toward more exclusive final-state measurements, the flexibility of the Ghent framework to handle inclusive, semi-inclusive, and exclusive channels within a unified formalism will be a crucial asset in the quest for a complete understanding of the lepton-nucleus interaction across the vector and axial sectors.

\begin{acknowledgments}
    We wish to express our sincere gratitude to the many colleagues and collaborators, especially J.T.~Sobczyk, whose insights and support have been instrumental to this work over the past several years. We would like to specifically thank M.~Martini for his invaluable discussions regarding the interferences. This work was partially supported by the Ghent University Special Research Fund (BOF) and the Research Foundation--Flanders (FWO). R.G.-J. was supported by projects RYC2022-035203-I funded by MCIN/AEI/10.13039/501100011033/FEDER and FSE+, UE; and by “Ayudas para Atracción de Investigadores con Alto Potencial-modalidad A” funded by VII PPIT-US. A.N. was supported by the Neutrino Theory Network (NTN) under Award Number DE-AC02-07CH11359.
\end{acknowledgments}


\bibliographystyle{apsrev4-2.bst}
\bibliography{bibliography}

@article{Mott:1929,
    author = {Mott, N. F. and Bohr, N.},
    title = {The scattering of fast electrons by atomic nuclei},
    doi = {10.1098/rspa.1929.0127},
    journal = {Proc. R. Soc. Lond. A},
    volume = {124},
    pages = {425--442},
    year = {1929}
}

@article{NuSTEC:2017hzk,
    author = "Alvarez-Ruso, L. and others",
    collaboration = "NuSTEC",
    title = "{NuSTEC White Paper: Status and challenges of neutrino\textendash{}nucleus scattering}",
    doi = "10.1016/j.ppnp.2018.01.006",
    journal = "Prog. Part. Nucl. Phys.",
    volume = "100",
    pages = "1--68",
    year = "2018"
}

@article{T2K:2011qtm,
    author = "Abe, K. and others",
    collaboration = "T2K",
    title = "{The T2K Experiment}",
    doi = "10.1016/j.nima.2011.06.067",
    journal = "Nucl. Instrum. Meth. A",
    volume = "659",
    pages = "106--135",
    year = "2011"
}

@article{NOvA:2016vij,
    author = "Adamson, P. and others",
    collaboration = "NOvA",
    title = "{First measurement of muon-neutrino disappearance in NOvA}",
    doi = "10.1103/PhysRevD.93.051104",
    journal = "Phys. Rev. D",
    volume = "93",
    number = "5",
    pages = "051104",
    year = "2016"
}

@article{T2K:2019bcf,
    author = "Abe, K. and others",
    collaboration = "T2K",
    title = "{Constraint on the matter\textendash{}antimatter symmetry-violating phase in neutrino oscillations}",
    doi = "10.1038/s41586-020-2177-0",
    journal = "Nature",
    volume = "580",
    number = "7803",
    pages = "339--344",
    year = "2020",
    note = "[Erratum: Nature 583, E16 (2020)]"
}

@misc{Hyper-Kamiokande:2018ofw,
    author = "Abe, K. and others",
    collaboration = "Hyper-Kamiokande",
    title = "{Hyper-Kamiokande Design Report}",
    eprint = "1805.04163",
    archivePrefix = "arXiv",
    primaryClass = "physics.ins-det",
    month = "5",
    year = "2018"
}

@article{DUNE:2020lwj,
    author = "Abi, Babak and others",
    collaboration = "DUNE",
    title = "{Deep Underground Neutrino Experiment (DUNE), Far Detector Technical Design Report, Volume I Introduction to DUNE}",
    doi = "10.1088/1748-0221/15/08/T08008",
    journal = "JINST",
    volume = "15",
    number = "08",
    pages = "T08008",
    year = "2020"
}

@article{Amaro:2019zos,
    author = "Amaro, J. E. and Barbaro, M. B. and Caballero, J. A. and Gonz\'alez-Jim\'enez, R. and Megias, G. D. and Ruiz Simo, I.",
    title = "{Electron- versus neutrino-nucleus scattering}",
    eprint = "1912.10612",
    archivePrefix = "arXiv",
    primaryClass = "nucl-th",
    doi = "10.1088/1361-6471/abb128",
    journal = "J. Phys. G",
    volume = "47",
    number = "12",
    pages = "124001",
    year = "2020"
}

@article{Lovato:2023raf,
    author = "Lovato, Alessandro and Nikolakopoulos, Alexis and Rocco, Noemi and Steinberg, Noah",
    title = "{Lepton-Nucleus Interactions within Microscopic Approaches}",
    eprint = "2308.00736",
    archivePrefix = "arXiv",
    primaryClass = "nucl-th",
    reportNumber = "FERMILAB-PUB-23-388-T",
    doi = "10.3390/universe9080367",
    journal = "Universe",
    volume = "9",
    number = "8",
    pages = "367",
    year = "2023"
}

@book{Walecka:1995mi,
    author = "Walecka, J. D.",
    title = "{Theoretical nuclear and subnuclear physics}",
    volume = "16",
    year = "1995",
   publisher     = "Oxford University Press",
}

@article{Benhar:1994hw,
    author = "Benhar, O. and Fabrocini, A. and Fantoni, S. and Sick, I.",
    title = "{Spectral function of finite nuclei and scattering of GeV electrons}",
    doi = "10.1016/0375-9474(94)90920-2",
    journal = "Nucl. Phys. A",
    volume = "579",
    pages = "493--517",
    year = "1994"
}

@article{Gonzalez-Jimenez:2019ejf,
    author = "Gonz\'alez-Jim\'enez, R. and Barbaro, M. B. and Caballero, J. A. and Donnelly, T. W. and Jachowicz, N. and Megias, G. D. and Niewczas, K. and Nikolakopoulos, A. and Ud\'\i{}as, J. M.",
    title = "{Constraints in modeling the quasielastic response in inclusive lepton-nucleus scattering}",
    eprint = "1909.07497",
    archivePrefix = "arXiv",
    primaryClass = "nucl-th",
    doi = "10.1103/PhysRevC.101.015503",
    journal = "Phys. Rev. C",
    volume = "101",
    number = "1",
    pages = "015503",
    year = "2020"
}

@article{Frullani:1984nn,
    author = "Frullani, S. and Mougey, J.",
    title = "{Single Particle Properties of Nuclei Through (e, e' p) Reactions}",
    journal = "Adv. Nucl. Phys.",
    volume = "14",
    pages = "1--283",
    year = "1984"
}

@article{Moniz:1971mt,
    author = "Moniz, E. J. and Sick, I. and Whitney, R. R. and Ficenec, J. R. and Kephart, Robert D. and Trower, W. P.",
    title = "{Nuclear Fermi momenta from quasielastic electron scattering}",
    doi = "10.1103/PhysRevLett.26.445",
    journal = "Phys. Rev. Lett.",
    volume = "26",
    pages = "445--448",
    year = "1971"
}

@article{Donnelly:1970rk,
    author = "Donnelly, T. W.",
    title = "{Quasielastic electron scattering in a square-well shell model}",
    doi = "10.1016/0375-9474(70)90959-0",
    journal = "Nucl. Phys. A",
    volume = "150",
    pages = "393--416",
    year = "1970"
}

@book{Boffi:1996ikg,
    author = "Boffi, Sigfrido and Giusti, Carlotta and Pacati, Franco davide and Radici, Marco",
    title = "{Electromagnetic Response of Atomic Nuclei}",
    isbn = "978-0-19-851774-0",
    publisher = "Clarendon Press",
    address = "Oxford UK",
    series = "Oxford Studies in Nuclear Physics",
    volume = "20",
    year = "1996"
}

@article{Benhar:2006wy,
    author = "Benhar, O. and Day, D. and Sick, I.",
    title = "{Inclusive quasi-elastic electron-nucleus scattering}",
    doi = "10.1103/RevModPhys.80.189",
    journal = "Rev. Mod. Phys.",
    volume = "80",
    pages = "189--224",
    year = "2008"
}

@article{Barreau:1983ht,
    author = "Barreau, P. and others",
    title = "{Deep Inelastic electron Scattering from Carbon}",
    doi = "10.1016/0375-9474(83)90217-8",
    journal = "Nucl. Phys. A",
    volume = "402",
    pages = "515--540",
    year = "1983"
}

@article{Jourdan:1996np,
    author = "Jourdan, J.",
    title = "{Quasielastic response functions: The Coulomb sum revisited}",
    doi = "10.1016/0375-9474(96)00143-1",
    journal = "Nucl. Phys. A",
    volume = "603",
    pages = "117--160",
    year = "1996"
}

@article{CZYZ196347,
    author = {Wieslaw Czyż and Kurt Gottfried},
    title = {Inelastic electron scattering from fluctuations in the nuclear charge distribution},
    doi = {10.1016/0003-4916(63)90224-0},
    journal = {Annals of Physics},
    volume = {21},
    number = {1},
    pages = {47-71},
    year = {1963}
}

@article{Noble:1980my,
    author = "Noble, J. V.",
    title = "{Modification of the nucleon's properties in nuclear matter}",
    doi = "10.1103/PhysRevLett.46.412",
    journal = "Phys. Rev. Lett.",
    volume = "46",
    pages = "412--415",
    year = "1981"
}

@article{DeForest:1983ahx,
    author = "De Forest, T.",
    title = "{Off-Shell Electron-Nucleon Cross-Sections. The Impulse Approximation}",
    doi = "10.1016/0375-9474(83)90124-0",
    journal = "Nucl. Phys. A",
    volume = "392",
    pages = "232--248",
    year = "1983"
}

@article{Celenza:1986vq,
    author = "Celenza, L. S. and Harindranath, A. and Shakin, C. M.",
    title = "{Sum Rules for the Longitudinal Response in Inclusive Electron Scattering}",
    doi = "10.1103/PhysRevC.33.1012",
    journal = "Phys. Rev. C",
    volume = "33",
    pages = "1012--1019",
    year = "1986"
}

@article{Mulders:1987ka,
    author = "Mulders, P. J.",
    editor = "Homma, S. and Yamazaki, T. and Morita, M. and Nakai, K.",
    title = "{Modifications os Nucleons in Nuclei in Quasielastic Electron-Nucleus Scattering}",
    reportNumber = "NIKHEF-P-3",
    doi = "10.1016/0375-9474(88)90829-9",
    journal = "Nucl. Phys. A",
    volume = "478",
    pages = "49C--66C",
    year = "1988"
}

@article{Chinn:1988ky,
    author = "Chinn, C. R. and Picklesimer, A. and Van Orden, J. W.",
    title = "{Final State Interactions and Relativistic Effects in the Quasielastic ($e$, $e^\prime$) Reaction}",
    reportNumber = "CEBAF-PR-88-020, LA-UR-88-4109, CEBAF PR 88 020, DOE-ER-40150-57",
    doi = "10.1103/PhysRevC.40.790",
    journal = "Phys. Rev. C",
    volume = "40",
    pages = "790--812",
    year = "1989"
}

@article{Capuzzi:1991qd,
    author = "Capuzzi, F. and Giusti, C. and Pacati, F. D.",
    title = "{Final state interaction in electromagnetic response functions}",
    doi = "10.1016/0375-9474(91)90269-C",
    journal = "Nucl. Phys. A",
    volume = "524",
    pages = "681--705",
    year = "1991"
}

@article{Boffi:1991hx,
    author = "Boffi, S. and Radici, M.",
    title = "{Nuclear response in electromagnetic quasifree knockout including two-body currents}",
    doi = "10.1016/0375-9474(91)90435-9",
    journal = "Nucl. Phys. A",
    volume = "526",
    pages = "602--622",
    year = "1991"
}

@article{Alberico:1987zz,
    author = "Alberico, W. M. and Czerski, P. and Ericson, M. and Molinari, A.",
    title = "{Inclusive charge longitudinal response in finite nuclei}",
    doi = "10.1016/0375-9474(87)90548-3",
    journal = "Nucl. Phys. A",
    volume = "462",
    pages = "269--289",
    year = "1987"
}

@article{Brieva:1987zz,
    author = "Brieva, F. A. and Dellafiore, Alberto",
    title = "{Random phase approximation effects in the longitudinal response function of C-12}",
    doi = "10.1103/PhysRevC.36.899",
    journal = "Phys. Rev. C",
    volume = "36",
    pages = "899--910",
    year = "1987"
}

@article{Co:1988zku,
    author = "Co', G. and Quader, K. F. and Smith, R. D. and Wambach, J.",
    title = "{Charge response in 12 C and 40 Ca}",
    doi = "10.1016/0375-9474(88)90522-2",
    journal = "Nucl. Phys. A",
    volume = "485",
    pages = "61--84",
    year = "1988"
}

@article{MARIASARUIS199357,
    author = {Anna {Maria Saruis}},
    title = {Self-consistent HF-RPA description of electron and photon nuclear reactions with Skyrme forces},
    doi = {10.1016/0370-1573(93)90123-U},
    journal = {Physics Reports},
    volume = {235},
    number="2",
    pages = {57-188},
    year = {1993}
}

@article{Bauer:2003yv,
    author = "Bauer, E.",
    title = "{The Role of the Delta(1232) on the longitudinal response in the inclusive electron scattering reaction}",
    eprint = "nucl-th/0302074",
    archivePrefix = "arXiv",
    doi = "10.1016/S0375-9474(03)01577-X",
    journal = "Nucl. Phys. A",
    volume = "725",
    pages = "149--170",
    year = "2003"
}

@article{Pandey:2014tza,
    author = "Pandey, V. and Jachowicz, N. and Van Cuyck, T. and Ryckebusch, J. and Martini, M.",
    title = "{Low-energy excitations and quasielastic contribution to electron-nucleus and neutrino-nucleus scattering in the continuum random-phase approximation}",
    eprint = "1412.4624",
    archivePrefix = "arXiv",
    primaryClass = "nucl-th",
    doi = "10.1103/PhysRevC.92.024606",
    journal = "Phys. Rev. C",
    volume = "92",
    number = "2",
    pages = "024606",
    year = "2015"
}

@article{Jachowicz:2021ieb,
    author = "Jachowicz, Natalie and Nikolakopoulos, Alexis",
    title = "{Nuclear medium effects in neutrino- and antineutrino-nucleus scattering}",
    eprint = "2110.11321",
    archivePrefix = "arXiv",
    primaryClass = "nucl-th",
    doi = "10.1140/epjs/s11734-021-00286-8",
    journal = "Eur. Phys. J. ST",
    volume = "230",
    number = "24",
    pages = "4339--4356",
    year = "2021"
}

@article{Wehrberger:1987ri,
    author = "Wehrberger, K. and Beck, F.",
    title = "{Relativistic Random Phase Approximation Response Function for Quasielastic Electron Scattering in Local Density Approximation}",
    doi = "10.1103/PhysRevC.35.298",
    journal = "Phys. Rev. C",
    volume = "35",
    pages = "298--304",
    year = "1987",
    note = "[Erratum: Phys.Rev.C 35, 2337--2337 (1987)]"
}

@article{Horowitz:1990bm,
    author = "Horowitz, C. J. and Piekarewicz, J.",
    title = "{Nuclear response functions in quasielastic electron scattering}",
    doi = "10.1016/0375-9474(90)90105-U",
    journal = "Nucl. Phys. A",
    volume = "511",
    pages = "461--486",
    year = "1990"
}

@article{VanderSluys:1995rp,
    author = "Van der Sluys, V. and Ryckebusch, J. and Waroquier, M.",
    title = "{Two-body currents in inclusive electron scattering}",
    eprint = "nucl-th/9503008",
    archivePrefix = "arXiv",
    reportNumber = "SSF-94-04-03, SSF94-04-03",
    doi = "10.1103/PhysRevC.51.2664",
    journal = "Phys. Rev. C",
    volume = "51",
    pages = "2664--2670",
    year = "1995"
}

@article{Lovato:2016gkq,
    author = "Lovato, A. and Gandolfi, S. and Carlson, J. and Pieper, Steven C. and Schiavilla, R.",
    title = "{Electromagnetic response of $^{12}$C: A first-principles calculation}",
    eprint = "1605.00248",
    archivePrefix = "arXiv",
    primaryClass = "nucl-th",
    reportNumber = "LA-UR-16-22606, JLAB-THY-16-2241",
    doi = "10.1103/PhysRevLett.117.082501",
    journal = "Phys. Rev. Lett.",
    volume = "117",
    number = "8",
    pages = "082501",
    year = "2016"
}

@article{Pastore:2019urn,
    author = "Pastore, Saori and Carlson, Joseph and Gandolfi, Stefano and Schiavilla, Rocco and Wiringa, Robert B.",
    title = "{Quasielastic lepton scattering and back-to-back nucleons in the short-time approximation}",
    eprint = "1909.06400",
    archivePrefix = "arXiv",
    primaryClass = "nucl-th",
    reportNumber = "LA-UR-19-29015",
    doi = "10.1103/PhysRevC.101.044612",
    journal = "Phys. Rev. C",
    volume = "101",
    number = "4",
    pages = "044612",
    year = "2020"
}

@article{Villars:1947,
    author = "Villars, Felix",
    title = "{The Magnetic Exchange Moments for H$_3$ and He$_3$}",
    journal = "Helv. Phys. Acta",
    volume = "20",
    pages = "476--490",
    year = "1947"
}

@article{Kohno:1981dg,
    author = "Kohno, M. and Ohtsuka, N.",
    title = "{MESON EXCHANGE CURRENT CONTRIBUTION TO QUASIFREE ELECTRON SCATTERING}",
    doi = "10.1016/0370-2693(81)90919-9",
    journal = "Phys. Lett. B",
    volume = "98",
    pages = "335--339",
    year = "1981"
}

@article{Fabrocini:1996bu,
    author = "Fabrocini, Adelchi",
    title = "{The Inclusive transverse response of nuclear matter}",
    eprint = "nucl-th/9609027",
    archivePrefix = "arXiv",
    reportNumber = "IFUP-TH-57-96",
    doi = "10.1103/PhysRevC.55.338",
    journal = "Phys. Rev. C",
    volume = "55",
    pages = "338--348",
    year = "1997"
}

@article{Alberico:1989aja,
    author = "Alberico, W. M. and Donnelly, T. W. and Molinari, A.",
    title = "{Pionic Effects in Quasielastic Electron Scattering}",
    reportNumber = "MIT-CTP-1794",
    doi = "10.1016/0375-9474(90)93832-Q",
    journal = "Nucl. Phys. A",
    volume = "512",
    pages = "541--590",
    year = "1990"
}

@article{Amaro:1994fx,
    author = "Amaro, Jose Enrique and Lallena, A. M. and Co, G.",
    title = "{Meson exchange currents in quasielastic electron scattering from C-12 and Ca-40 nuclei}",
    doi = "10.1016/0375-9474(94)90752-8",
    journal = "Nucl. Phys. A",
    volume = "578",
    pages = "365--396",
    year = "1994"
}

@article{Amaro:2001xz,
    author = "Amaro, Jose Enrique and Barbaro, M. B. and Caballero, J. A. and Donnelly, T. W. and Molinari, A.",
    title = "{Relativistic pionic effects in quasielastic electron scattering}",
    eprint = "nucl-th/0106035",
    archivePrefix = "arXiv",
    reportNumber = "MIT-CTP-3127",
    doi = "10.1016/S0375-9474(01)01253-2",
    journal = "Nucl. Phys. A",
    volume = "697",
    pages = "388--428",
    year = "2002"
}

@article{Amaro:2003yd,
    author = "Amaro, Jose Enrique and Barbaro, M. B. and Caballero, J. A. and Donnelly, T. W. and Molinari, A.",
    title = "{Delta isobar relativistic meson exchange currents in quasielastic electron scattering}",
    eprint = "nucl-th/0301023",
    archivePrefix = "arXiv",
    doi = "10.1016/S0375-9474(03)01269-7",
    journal = "Nucl. Phys. A",
    volume = "723",
    pages = "181--204",
    year = "2003"
}

@article{Casale:2023hyz,
    author = "Casale, Paloma Rodriguez and Amaro, Jose Enrique and Barbaro, Maria B.",
    title = "{Meson-Exchange Currents in Quasielastic Electron Scattering in a Generalized Superscaling Approach}",
    eprint = "2307.15783",
    archivePrefix = "arXiv",
    primaryClass = "nucl-th",
    doi = "10.3390/sym15091709",
    journal = "Symmetry",
    volume = "15",
    number = "9",
    pages = "1709",
    year = "2023"
}

@article{Dekker:1992px,
    author = "Dekker, M. J. and Brussaard, P. J. and Tjon, J. A.",
    title = "{Two-body currents and the transverse response of nuclei at large momentum transfer}",
    doi = "10.1016/0370-2693(92)91215-U",
    journal = "Phys. Lett. B",
    volume = "289",
    pages = "255--260",
    year = "1992"
}

@article{Franco-Munoz:2023zoa,
    author = "Franco-Munoz, T. and Garc\'\i{}a-Marcos, J. and Gonz\'alez-Jim\'enez, R. and Ud\'\i{}as, J. M.",
    title = "{Relativistic two-body currents for one-nucleon knockout in electron-nucleus scattering}",
    eprint = "2306.10823",
    archivePrefix = "arXiv",
    primaryClass = "nucl-th",
    doi = "10.1103/PhysRevC.108.064608",
    journal = "Phys. Rev. C",
    volume = "108",
    number = "6",
    pages = "064608",
    year = "2023"
}

@article{Franco-Munoz:2022jcl,
    author = "Franco-Munoz, T. and Gonz{\'a}lez-Jim{\'e}nez, R. and Ud{\'\i}as, J. M.",
    title = "{Effects of two-body currents in the one-particle one-hole electromagnetic responses within a relativistic model}",
    eprint = "2203.09996",
    archivePrefix = "arXiv",
    primaryClass = "nucl-th",
    doi = "10.1088/1361-6471/ad9eca",
    journal = "J. Phys. G",
    volume = "52",
    number = "2",
    pages = "025103",
    year = "2025"
}

@article{Lovato:2023khk,
    author = "Lovato, Alessandro and Rocco, Noemi and Steinberg, Noah",
    title = "{One- and two-body current contributions to lepton-nucleus scattering}",
    eprint = "2312.12545",
    archivePrefix = "arXiv",
    primaryClass = "nucl-th",
    reportNumber = "FERMILAB-PUB-23-0823-T",
    doi = "10.1103/m645-5whh",
    journal = "Phys. Rev. C",
    volume = "112",
    number = "4",
    pages = "045501",
    year = "2025"
}

@article{Carlson:2001mp,
    author = "Carlson, J. and Jourdan, J. and Schiavilla, R. and Sick, I.",
    title = "{Longitudinal and transverse quasielastic response functions of light nuclei}",
    eprint = "nucl-th/0106047",
    archivePrefix = "arXiv",
    reportNumber = "LA-UR-01-3235, JLAB-THY-01-19",
    doi = "10.1103/PhysRevC.65.024002",
    journal = "Phys. Rev. C",
    volume = "65",
    pages = "024002",
    year = "2002"
}

@misc{Franco-Munoz:2026hxi,
    author = "Franco-Munoz, T. and McKean, J. and Garc{\'\i}a-Marcos, J. and Hooft, M. and Gonz{\'a}lez-Jim{\'e}nez, R. and Jachowicz, N. and Ud{\'\i}as, J. M.",
    title = "{Two-body current and axial form factor effects in charged-current quasielastic neutrino-nucleus scattering within the NEUT event generator}",
    eprint = "2605.00756",
    archivePrefix = "arXiv",
    primaryClass = "nucl-th",
    month = "5",
    year = "2026"
}

@article{Towner:1987zz,
    author = "Towner, I. S.",
    title = "{Quenching of spin matrix elements in nuclei}",
    doi = "10.1016/0370-1573(87)90138-4",
    journal = "Phys. Rept.",
    volume = "155",
    pages = "263--377",
    year = "1987"
}

@article{Alberico:1983zg,
    author = "Alberico, W. M. and Ericson, Magda and Molinari, A.",
    title = "{The Role of Two Particles - Two Holes Excitations in the Spin - Isospin Nuclear Response}",
    reportNumber = "CERN-TH-3635",
    doi = "10.1016/0003-4916(84)90155-6",
    journal = "Annals Phys.",
    volume = "154",
    pages = "356",
    year = "1984"
}

@article{Martini:2011ui,
    author = "Martini, M.",
    editor = "Blondel, Alain and Efthymiopoulos, Ilias and Prior, Gersende",
    title = "{Two Particle-Two Hole Excitations in Charged Current Quasielastic Neutrino-Nucleus Interactions}",
    eprint = "1110.5895",
    archivePrefix = "arXiv",
    primaryClass = "hep-ph",
    doi = "10.1088/1742-6596/408/1/012041",
    journal = "J. Phys. Conf. Ser.",
    volume = "408",
    pages = "012041",
    year = "2013"
}

@article{Dekker:1994yc,
    author = "Dekker, M. J. and Brussaard, P. J. and Tjon, J. A.",
    title = "{Relativistic meson exchange and isobar currents in electron scattering: Noninteracting Fermi gas analysis}",
    doi = "10.1103/PhysRevC.49.2650",
    journal = "Phys. Rev. C",
    volume = "49",
    pages = "2650--2670",
    year = "1994"
}

@article{Dekker:1991ph,
    author = "Dekker, M. J. and Brussaard, P. J. and Tjon, J. A.",
    title = "{Relativistic meson exchange currents in the dip region}",
    doi = "10.1016/0370-2693(91)91034-S",
    journal = "Phys. Lett. B",
    volume = "266",
    pages = "249--254",
    year = "1991"
}

@article{Gil:1997bm,
    author = "Gil, A. and Nieves, J. and Oset, E.",
    title = "{Many body approach to the inclusive (e, e-prime) reaction from the quasielastic to the Delta excitation region}",
    eprint = "nucl-th/9711009",
    archivePrefix = "arXiv",
    reportNumber = "UG-DFM-2-97, FTUV-97-18, IFIC-97-18",
    doi = "10.1016/S0375-9474(97)00513-7",
    journal = "Nucl. Phys. A",
    volume = "627",
    pages = "543--598",
    year = "1997"
}

@article{DePace:2003spn,
    author = "De Pace, A. and Nardi, M. and Alberico, W. M. and Donnelly, T. W. and Molinari, A.",
    title = "{The 2p - 2h electromagnetic response in the quasielastic peak and beyond}",
    eprint = "nucl-th/0304084",
    archivePrefix = "arXiv",
    doi = "10.1016/S0375-9474(03)01625-7",
    journal = "Nucl. Phys. A",
    volume = "726",
    pages = "303--326",
    year = "2003"
}

@article{RuizSimo:2016rtu,
    author = "Ruiz Simo, I. and Amaro, J. E. and Barbaro, M. B. and De Pace, A. and Caballero, J. A. and Donnelly, T. W.",
    title = "{Relativistic model of 2p-2h meson exchange currents in (anti)neutrino scattering}",
    eprint = "1604.08423",
    archivePrefix = "arXiv",
    primaryClass = "nucl-th",
    doi = "10.1088/1361-6471/aa6a06",
    journal = "J. Phys. G",
    volume = "44",
    number = "6",
    pages = "065105",
    year = "2017"
}

@article{Megias:2016lke,
    author = "Megias, G. D. and Amaro, J. E. and Barbaro, M. B. and Caballero, J. A. and Donnelly, T. W.",
    title = "{Inclusive electron scattering within the SuSAv2 meson-exchange current approach}",
    eprint = "1603.08396",
    archivePrefix = "arXiv",
    primaryClass = "nucl-th",
    doi = "10.1103/PhysRevD.94.013012",
    journal = "Phys. Rev. D",
    volume = "94",
    pages = "013012",
    year = "2016"
}

@phdthesis{Belocchi:2025lqp,
    author = "Belocchi, Valerio",
    title = "{Meson-exchange currents in lepton-nucleus scattering}",
    school = "Universit{\`a} degli Studi di Torino, Italy",
    year = "2025"
}

@article{Benhar:2015ula,
    author = "Benhar, Omar and Lovato, Alessandro and Rocco, Noemi",
    title = "{Contribution of two-particle\textendash{}two-hole final states to the nuclear response}",
    eprint = "1502.00887",
    archivePrefix = "arXiv",
    primaryClass = "nucl-th",
    doi = "10.1103/PhysRevC.92.024602",
    journal = "Phys. Rev. C",
    volume = "92",
    number = "2",
    pages = "024602",
    year = "2015"
}

@article{Donnelly:1978xa,
    author = "Donnelly, T. W. and Van Orden, J. W. and De Forest, Jr., T. and Hermans, W. C.",
    title = "{Meson Exchange Currents in Deep Inelastic Electron Scattering From Nuclei}",
    reportNumber = "ITP-594-STANFORD",
    doi = "10.1016/0370-2693(78)90890-0",
    journal = "Phys. Lett. B",
    volume = "76",
    pages = "393--396",
    year = "1978"
}

@article{Ryckebusch:1988bpw,
    author = "Ryckebusch, J. and Waroquier, M. and Heyde, K. and Moreau, J. and Ryckbosch, D.",
    title = "{An RPA model for the description of one-nucleon emission processes and application to 16 O(\ensuremath{\gamma}, N) reactions}",
    doi = "10.1016/0375-9474(88)90483-6",
    journal = "Nucl. Phys. A",
    volume = "476",
    pages = "237--271",
    year = "1988"
}

@article{Jachowicz:2002rr,
    author = "Jachowicz, N. and Heyde, K. and Ryckebusch, J. and Rombouts, S.",
    title = "{Continuum random phase approximation approach to charged current neutrino nucleus scattering}",
    doi = "10.1103/PhysRevC.65.025501",
    journal = "Phys. Rev. C",
    volume = "65",
    pages = "025501",
    year = "2002"
}

@article{Ryckebusch:1997gn,
    author = "Ryckebusch, J. and Van der Sluys, V. and Heyde, K. and Holvoet, H. and Van Nespen, W. and Waroquier, M. and Vanderhaeghen, M.",
    title = "{Electroinduced two nucleon knockout and correlations in nuclei}",
    eprint = "nucl-th/9702049",
    archivePrefix = "arXiv",
    reportNumber = "SSF-96-12-01",
    doi = "10.1016/S0375-9474(97)00385-0",
    journal = "Nucl. Phys. A",
    volume = "624",
    pages = "581--622",
    year = "1997"
}

@article{Janssen:1999xy,
    author = "Janssen, Stijn and Ryckebusch, Jan and Van Nespen, Wim and Debruyne, Dimitri",
    title = "{Spin dependent correlations and the semiexclusive O-16(e, e-prime p) reaction}",
    eprint = "nucl-th/9911054",
    archivePrefix = "arXiv",
    doi = "10.1016/S0375-9474(00)00058-0",
    journal = "Nucl. Phys. A",
    volume = "672",
    pages = "285--309",
    year = "2000"
}

@article{Ryckebusch:1993tf,
    author = "Ryckebusch, J. and Vanderhaeghen, M. and Machenil, L. and Waroquier, M.",
    title = "{Effects of the final state interaction in (gamma, p n) and (gamma, p p) processes}",
    eprint = "nucl-th/9307004",
    archivePrefix = "arXiv",
    reportNumber = "INWO-701",
    doi = "10.1016/0375-9474(94)90362-X",
    journal = "Nucl. Phys. A",
    volume = "568",
    pages = "828--854",
    year = "1994"
}

@article{VanCuyck:2016fab,
    author = "Van Cuyck, Tom and Jachowicz, Natalie and Gonz\'alez-Jim\'enez, Ra\'ul and Martini, Marco and Pandey, Vishvas and Ryckebusch, Jan and Van Dessel, Nils",
    title = "{Influence of short-range correlations in neutrino-nucleus scattering}",
    eprint = "1606.00273",
    archivePrefix = "arXiv",
    primaryClass = "nucl-th",
    doi = "10.1103/PhysRevC.94.024611",
    journal = "Phys. Rev. C",
    volume = "94",
    number = "2",
    pages = "024611",
    year = "2016"
}

@article{Ryckebusch:1999gu,
    author = "Ryckebusch, Jan and Debruyne, Dimitri and Van Nespen, Wim and Janssen, Stijn",
    title = "{Meson and isobar degrees of freedom in A (polarized e, e-prime polarized p) reactions at 0.2 less than or equal to Q**2 less than or equal to 0.8(GeV/c)**2}",
    eprint = "nucl-th/9904011",
    archivePrefix = "arXiv",
    doi = "10.1103/PhysRevC.60.034604",
    journal = "Phys. Rev. C",
    volume = "60",
    pages = "034604",
    year = "1999"
}

@article{VanCuyck:2017wfn,
    author = "Van Cuyck, T. and Jachowicz, N. and Gonz\'alez-Jim\'enez, R. and Ryckebusch, J. and Van Dessel, N.",
    title = "{Seagull and pion-in-flight currents in neutrino-induced $1N$ and $2N$ knockout}",
    eprint = "1702.06402",
    archivePrefix = "arXiv",
    primaryClass = "nucl-th",
    doi = "10.1103/PhysRevC.95.054611",
    journal = "Phys. Rev. C",
    volume = "95",
    number = "5",
    pages = "054611",
    year = "2017"
}

@article{Waroquier:1983fqb,
    author = "Waroquier, M. and Wenes, G. and Heyde, K.",
    title = "{An effective Skyrme-type interaction for nuclear structure calculations}",
    doi = "10.1016/0375-9474(83)90551-1",
    journal = "Nucl. Phys. A",
    volume = "404",
    pages = "298--332",
    year = "1983"
}

@article{Vautherin:1971aw,
    author = "Vautherin, D. and Brink, D. M.",
    title = "{Hartree-Fock calculations with Skyrme's interaction. 1. Spherical nuclei}",
    doi = "10.1103/PhysRevC.5.626",
    journal = "Phys. Rev. C",
    volume = "5",
    pages = "626--647",
    year = "1972"
}

@phdthesis{Jachowicz:2000phd,
    author = "Jachowicz, Natalie",
    title = "{Many-body description of neutrino-nucleus interactions}",
    school = "Ghent University",
    year = "2000"
}

@article{Amaro:1993306,
    author = {J.E. Amaro and G. Co and A.M. Lallena},
    title = {Electromagnetic Quasi-Elastic Responses in 12C},
    doi = {https://doi.org/10.1006/aphy.1993.1013},
    journal = {Annals of Physics},
    volume = {221},
    number = {2},
    pages = {306-340},
    year = {1993}
}

@phdthesis{Niewczas:2023phd,
    author = "Niewczas, Kajetan",
    title = "{Multinucleon knock-out in neutrino-nucleus scattering: merging theory and Monte Carlo simulations}",
    school = "University of Wrocław and Ghent University",
    year = "2023"
}

@book{Bjorken:1965sts,
    author = "Bjorken, James D. and Drell, Sidney D.",
    title = "{Relativistic Quantum Mechanics}",
    isbn = "978-0-07-005493-6",
    publisher = "McGraw-Hill",
    address = "New York",
    series = "International Series In Pure and Applied Physics",
    year = "1965"
}

@article{Galster:1971kv,
    author = "Galster, S. and Klein, H. and Moritz, J. and Schmidt, K. H. and Wegener, D. and Bleckwenn, J.",
    title = "{Elastic electron-deuteron scattering and the electric neutron form factor at four-momentum transfers 5fm$^{-2} < q^2 < 14$fm$^{-2}$}",
    reportNumber = "DESY-71-7",
    doi = "10.1016/0550-3213(71)90068-X",
    journal = "Nucl. Phys. B",
    volume = "32",
    pages = "221--237",
    year = "1971"
}

@article{Friar:1973hxj,
    author = "Friar, J. L.",
    title = "{Relativistic corrections to electron scattering by the deuteron and triton}",
    doi = "10.1016/0370-2693(73)90421-8",
    journal = "Phys. Lett. B",
    volume = "43",
    pages = "108--112",
    year = "1973"
}

@book{Ericson:1988gk,
    author = "Ericson, Torleif Erik Oskar and Weise, W.",
    title = "{Pions and Nuclei}",
    isbn = "978-0-19-852008-5",
    publisher = "Clarendon Press",
    address = "Oxford, UK",
    year = "1988"
}

@article{Ryckebusch:2019oya,
    author = "Ryckebusch, Jan and Cosyn, Wim and Vieijra, Tom and Casert, Corneel",
    title = "{Isospin composition of the high-momentum fluctuations in nuclei from asymptotic momentum distributions}",
    eprint = "1907.07259",
    archivePrefix = "arXiv",
    primaryClass = "nucl-th",
    doi = "10.1103/PhysRevC.100.054620",
    journal = "Phys. Rev. C",
    volume = "100",
    number = "5",
    pages = "054620",
    year = "2019"
}

@article{Mathiot:1989vw,
    author = "Mathiot, J. F.",
    title = "{Electromagnetic meson exchange currents at the nucleon mass scale}",
    doi = "10.1016/0370-1573(89)90079-3",
    journal = "Phys. Rept.",
    volume = "173",
    pages = "63--172",
    year = "1989"
}

@article{Mathiot:1984bs,
    author = "Mathiot, J. F.",
    title = "{WHAT DO WE LEARN FROM DEUTERON ELECTRODISINTEGRATION NEAR THRESHOLD AND THERMAL N P CAPTURE?}",
    doi = "10.1016/0375-9474(84)90679-1",
    journal = "Nucl. Phys. A",
    volume = "412",
    pages = "201--227",
    year = "1984"
}

@article{Amaro:2002mj,
    author = "Amaro, Jose Enrique and Barbaro, M. B. and Caballero, J. A. and Donnelly, T. W. and Molinari, A.",
    title = "{Gauge and Lorentz invariant one pion exchange currents in electron scattering from a relativistic Fermi gas}",
    eprint = "nucl-th/0204001",
    archivePrefix = "arXiv",
    reportNumber = "MIT-CTP-3167",
    doi = "10.1016/S0370-1573(02)00195-3",
    journal = "Phys. Rept.",
    volume = "368",
    pages = "317--407",
    year = "2002"
}

@article{Marcucci:1998tb,
    author = "Marcucci, L. E. and Riska, D. O. and Schiavilla, R.",
    title = "{Electromagnetic structure of the trinucleons}",
    eprint = "nucl-th/9805048",
    archivePrefix = "arXiv",
    reportNumber = "JLAB-THY-98-21",
    doi = "10.1103/PhysRevC.58.3069",
    journal = "Phys. Rev. C",
    volume = "58",
    pages = "3069--3084",
    year = "1998"
}

@article{Stoler:1993yk,
    author = "Stoler, P.",
    title = "{Baryon form-factors at high Q**2 and the transition to perturbative QCD}",
    doi = "10.1016/0370-1573(93)90088-U",
    journal = "Phys. Rept.",
    volume = "226",
    pages = "103--171",
    year = "1993"
}

@article{Frolov:1998pw,
    author = "Frolov, V. V. and others",
    title = "{Electroproduction of the Delta (1232) resonance at high momentum transfer}",
    eprint = "hep-ex/9808024",
    archivePrefix = "arXiv",
    reportNumber = "JLAB-PHY-98-08",
    doi = "10.1103/PhysRevLett.82.45",
    journal = "Phys. Rev. Lett.",
    volume = "82",
    pages = "45--48",
    year = "1999"
}

@article{Rarita:1941mf,
    author = "Rarita, William and Schwinger, Julian",
    title = "{On a theory of particles with half integral spin}",
    doi = "10.1103/PhysRev.60.61",
    journal = "Phys. Rev.",
    volume = "60",
    pages = "61",
    year = "1941"
}

@article{Vanderhaeghen:1995fe,
    author = "Vanderhaeghen, M. and Heyde, K. and Ryckebusch, J. and Waroquier, M.",
    title = "{Pion photoproduction through the Delta resonance region: Relativistic versus nonrelativistic unitary models}",
    doi = "10.1016/0375-9474(95)00386-1",
    journal = "Nucl. Phys. A",
    volume = "595",
    pages = "219--258",
    year = "1995"
}

@misc{Hooft:2026yyg,
    author = "Hooft, M. and Nikolakopoulos, A. and Garc{\'\i}a-Marcos, J. and De Backer, Y. and Franco-Munoz, T. and Niewczas, K. and Gonz{\'a}lez-Jim{\'e}nez, R. and Jachowicz, N.",
    title = "{Optimizing the description of the Delta region in the Ghent Hybrid model for single-pion production}",
    eprint = "2603.29486",
    archivePrefix = "arXiv",
    primaryClass = "nucl-th",
    month = "3",
    year = "2026"
}

@article{Chen:1988xq,
    author = "Chen, C. R. and Lee, T. S. H.",
    title = "{Excitation of the $\Delta$ Resonance in the $^{12}$C ($e$, $e^\prime$) Reaction}",
    doi = "10.1103/PhysRevC.38.2187",
    journal = "Phys. Rev. C",
    volume = "38",
    pages = "2187--2192",
    year = "1988"
}

@article{Oset:1987re,
    author = "Oset, E. and Salcedo, L. L.",
    title = "{$\Delta$ Selfenergy in Nuclear Matter}",
    doi = "10.1016/0375-9474(87)90185-0",
    journal = "Nucl. Phys. A",
    volume = "468",
    pages = "631--652",
    year = "1987"
}

@article{Nieves:1993ev,
    author = "Nieves, J. and Oset, E. and Garcia-Recio, C.",
    title = "{A Theoretical approach to pionic atoms and the problem of anomalies}",
    doi = "10.1016/0375-9474(93)90245-S",
    journal = "Nucl. Phys. A",
    volume = "554",
    pages = "509--553",
    year = "1993"
}

@article{Hjorth-Jensen:1993qiu,
    author = "Hjorth-Jensen, M. and Muther, H. and Polls, A.",
    title = "{Width of the Delta resonance in nuclei}",
    eprint = "nucl-th/9309019",
    archivePrefix = "arXiv",
    reportNumber = "TU-93-1609",
    doi = "10.1103/PhysRevC.50.501",
    journal = "Phys. Rev. C",
    volume = "50",
    pages = "501--504",
    year = "1994"
}

@article{Wilbois:1996zz,
    author = "Wilbois, Th. and Wilhelm, P. and Arenhovel, H.",
    title = "{Comment on ``Multinucleon mechanisms in (gamma, N) and (gamma, NN) reactions''}",
    doi = "10.1103/PhysRevC.54.3311",
    journal = "Phys. Rev. C",
    volume = "54",
    pages = "3311--3312",
    year = "1996"
}

@article{Gottfried:1958,
    author = "Gottfried, K.",
    title = "On the determination of the nuclear pair correlation function from the high energy photo-effect",
    journal = "Nucl. Phys.",
    volume = "5",
    pages = "5557",
    year = "1958"
}

@article{Megias:2018ujz,
    author = "Megias, G. D. and Barbaro, M. B. and Caballero, J. A. and Dolan, S.",
    title = "{Analysis of the MINERvA antineutrino double-differential cross sections within the SuSAv2 model including meson-exchange currents}",
    eprint = "1807.10532",
    archivePrefix = "arXiv",
    primaryClass = "nucl-th",
    doi = "10.1103/PhysRevD.99.113002",
    journal = "Phys. Rev. D",
    volume = "99",
    number = "11",
    pages = "113002",
    year = "2019"
}

@misc{TaniaPriv,
  author = {Franco-Munoz, T.},
  howpublished = {private communication},
  year = {2022}
}

@article{Tropiano:2021qgf,
    author = "Tropiano, A. J. and Bogner, S. K. and Furnstahl, R. J.",
    title = "{Short-range correlation physics at low renormalization group resolution}",
    eprint = "2105.13936",
    archivePrefix = "arXiv",
    primaryClass = "nucl-th",
    doi = "10.1103/PhysRevC.104.034311",
    journal = "Phys. Rev. C",
    volume = "104",
    number = "3",
    pages = "034311",
    year = "2021"
}

@phdthesis{Gearhart:1994phd,
    author = "Gearhart, C.",
    title = "Self-consistently dressed nucleons and nuclear wave functions",
    school = "Washington University",
    year = "1994"
}

@article{Pieper:1992gr,
    author = "Pieper, Steven C. and Wiringa, Robert B. and Pandharipande, V. R.",
    title = "{Variational calculation of the ground state of O-16}",
    doi = "10.1103/PhysRevC.46.1741",
    journal = "Phys. Rev. C",
    volume = "46",
    pages = "1741--1756",
    year = "1992"
}

@article{Reid:1968sq,
    author = "Reid, Jr., Roderick V.",
    title = "{Local phenomenological nucleon-nucleon potentials}",
    doi = "10.1016/0003-4916(68)90126-7",
    journal = "Annals Phys.",
    volume = "50",
    pages = "411--448",
    year = "1968"
}

@article{Wiringa:1984tg,
    author = "Wiringa, Robert B. and Smith, R. A. and Ainsworth, T. L.",
    title = "{Nucleon Nucleon Potentials with and Without Delta (1232) Degrees of Freedom}",
    doi = "10.1103/PhysRevC.29.1207",
    journal = "Phys. Rev. C",
    volume = "29",
    pages = "1207--1221",
    year = "1984"
}

@article{Schiavilla:1985gb,
    author = "Schiavilla, R. and Pandharipande, V. R. and Wiringa, Robert B.",
    title = "{Momentum distributions in a A = 3 and 4 nuclei}",
    doi = "10.1016/0375-9474(86)90003-5",
    journal = "Nucl. Phys. A",
    volume = "449",
    pages = "219--242",
    year = "1986"
}

@article{Blomqvist:1998gq,
    author = "Blomqvist, K. I. and others",
    title = "{Investigation of short range nucleon nucleon correlations using the reaction C-12(e,e' p p) in close to 4pi geometry}",
    doi = "10.1016/S0370-2693(98)00024-0",
    journal = "Phys. Lett. B",
    volume = "421",
    pages = "71--78",
    year = "1998"
}

@article{Wiringa:2013ala,
    author = "Wiringa, R. B. and Schiavilla, R. and Pieper, Steven C. and Carlson, J.",
    title = "{Nucleon and nucleon-pair momentum distributions in $A \le 12$ nuclei}",
    eprint = "1309.3794",
    archivePrefix = "arXiv",
    primaryClass = "nucl-th",
    reportNumber = "JLAB-THY-13-1789",
    doi = "10.1103/PhysRevC.89.024305",
    journal = "Phys. Rev. C",
    volume = "89",
    number = "2",
    pages = "024305",
    year = "2014"
}

@article{Lonardoni:2017egu,
    author = "Lonardoni, D. and Lovato, A. and Pieper, Steven C. and Wiringa, R. B.",
    title = "{Variational calculation of the ground state of closed-shell nuclei up to $A=40$}",
    eprint = "1705.04337",
    archivePrefix = "arXiv",
    primaryClass = "nucl-th",
    reportNumber = "LA-UR-17-23856",
    doi = "10.1103/PhysRevC.96.024326",
    journal = "Phys. Rev. C",
    volume = "96",
    number = "2",
    pages = "024326",
    year = "2017"
}

@article{Ryckebusch:1995usx,
    author = "Ryckebusch, J. and Vanderhaeghen, M. and Heyde, K. and Waroquier, M.",
    title = "{Short-range correlations in (e,e$'$p) and (e,e$'$pp) reactions on complex nuclei}",
    eprint = "nucl-th/9503007",
    archivePrefix = "arXiv",
    reportNumber = "SSF94-10-01",
    doi = "10.1016/0370-2693(95)00323-D",
    journal = "Phys. Lett. B",
    volume = "350",
    pages = "1--7",
    year = "1995"
}

@article{Shimizu:1980kb,
    author = "Shimizu, K. and Faessler, A.",
    title = "{THE ABSORPTIVE PION NUCLEUS OPTICAL POTENTIAL}",
    doi = "10.1016/0375-9474(80)90112-8",
    journal = "Nucl. Phys. A",
    volume = "333",
    pages = "495--513",
    year = "1980"
}

\end{document}